\documentclass[pre,superscriptaddress,showpacs,twocolumn,longbibliography]{revtex4-1}
\usepackage{color}
\usepackage[usenames,dvipsnames]{xcolor}
 
\usepackage{graphicx}
\usepackage{epsfig}
\usepackage{dcolumn}
\usepackage{bm}
\usepackage{mathrsfs}
\usepackage{multirow}
\usepackage[all]{xy}
\usepackage{pbox}
\usepackage{amssymb}
\usepackage{scalerel}

\usepackage{color}
\definecolor{myblue}{RGB}{65,105,225}
\definecolor{mygreen}{RGB}{34,139,34}
\definecolor{myorange}{RGB}{255,69,0}
\usepackage[colorlinks,linkcolor=myorange,urlcolor=myblue,citecolor=myblue]{hyperref}

\def\(({\left(}
\def\)){\right)}
\def\[[{\left[}
\def\]]{\right]}

\newcommand{\hD}{{\hat{D}}}
\newcommand{\hS}{{\hat{\sigma}}}

\newcommand{\be}{\begin{equation}}
\newcommand{\ee}{\end{equation}}
\newcommand{\ben}{\begin{eqnarray}}
\newcommand{\een}{\end{eqnarray}}
\newcommand{\beq}{\begin{equation}}
\newcommand{\eeq}{\end{equation}}

\newcommand{\la}{\langle}
\newcommand{\ra}{\rangle}

\newcommand{\vrr}{{\bf{r}}}
\newcommand{\vxi}{{\bm{\xi}}}
\newcommand{\vnabla}{{\bm{\nabla}}}
\newcommand{\vj}{{\bf{j}}}
\newcommand{\vQ}{{\bf{Q}}}
\newcommand{\vE}{{\bf{E}}}
\newcommand{\vq}{{\bf{q}}}
\newcommand{\vv}{{\bf{v}}}

\newcommand{\vlamb}{{\bm{\lambda}}}
\newcommand{\vz}{{\bf{z}}}
\newcommand{\mA}{{\hat{\cal A}}}

\begin{document}

\title{Order and symmetry-breaking in the fluctuations of driven systems} 

\author{N. Tiz\'on-Escamilla}
\affiliation{Departamento de Electromagnetismo y F\'{\i}sica de la Materia, and Institute Carlos I for Theoretical and Computational Physics, Universidad de Granada, Granada 18071, Spain}

\author{C. P\'erez-Espigares}
\affiliation{University of Modena and Reggio Emilia, via G. Campi 213/b, 41125 Modena, Italy}
\affiliation{School of Physics and Astronomy, University of Nottingham, Nottingham, NG7 2RD, UK}

\author{P.L. Garrido}
\affiliation{Departamento de Electromagnetismo y F\'{\i}sica de la Materia, and Institute Carlos I for Theoretical and Computational Physics, Universidad de Granada, Granada 18071, Spain}

\author{P.I. Hurtado}
\email[]{phurtado@onsager.ugr.es}
\affiliation{Departamento de Electromagnetismo y F\'{\i}sica de la Materia, and Institute Carlos I for Theoretical and Computational Physics, Universidad de Granada, Granada 18071, Spain}

\date{\today}

\begin{abstract}
Dynamical phase transitions (DPTs) in the space of trajectories are one of the most intriguing phenomena of nonequilibrium physics, but their nature in realistic high-dimensional systems remains puzzling. Here we observe for the first time a DPT in the current vector statistics of an archetypal two-dimensional ($2d$) driven diffusive system, and characterize its properties using macroscopic fluctuation theory. The complex interplay among the external field, anisotropy and vector currents in $2d$ leads to a rich phase diagram, with different symmetry-broken fluctuation phases separated by lines of $1^{\text{st}}$- and $2^{\text{nd}}$-order DPTs. Remarkably, different types of $1d$ order in the form of jammed density waves emerge to hinder transport for low-current fluctuations, revealing a connection between rare events and self-organized structures which enhance their probability.
\end{abstract}

\maketitle 

{\bf \emph{Introduction}}-- The theory of critical phenomena is a cornerstone of modern theoretical physics \cite{binney92a,zinn-justin02a}. Indeed, phase transitions of all sorts appear ubiquitously in most domains of physics, from cosmological scales to the quantum world of elementary particles. In a typical $2^{\text{nd}}$-order phase transition order emerges  \emph{continuously} at some critical point, as captured by an order parameter, signaling the spontaneous breaking of a symmetry and an associated non-analyticity of the relevant thermodynamic potential. Conversely, $1^{\text{st}}$-order transitions are characterized by an \emph{abrupt} jump in the order parameter and a coexistente of different phases \cite{binney92a,zinn-justin02a}. In recent years these ideas have been extended to the realm of fluctuations, where \emph{dynamical} phase transitions (i.e. in the space of \emph{trajectories}) have been identified in different systems, both classical \cite{bertini15a,derrida07a,bodineau05a,bertini06a,lecomte07b,bodineau08a,hurtado11a,perez-espigares13a,hurtado14a,vaikuntanathan14a,jack15a,shpielberg16a,zarfaty16a,baek17a,karevski17a} and quantum \cite{garrahan10a,ates12a,lesanovsky13a,carollo17a}. Important examples include glass formers \cite{garrahan07a,garrahan09a,hedges09a,chandler10a,pitard11a,speck12a,pinchaipat17a,abou17a}, micromasers and superconducting transistors \cite{garrahan11a,genway12a}, or applications such as DPT-based quantum thermal switches \cite{manzano14a,manzano16a,manzano17a}. 

DPTs appear when conditioning a system to have a fixed value of some time-integrated observable, as e.g. the current or the activity. The different dynamical phases correspond to different types of trajectories adopted by the system to sustain atypical values of this observable. Interestingly, some dynamical phases may display emergent order and collective rearrangements in their trajectories, including symmetry-breaking phenomena \cite{bodineau05a,hurtado11a,perez-espigares13a,hurtado14a}, while the \emph{large deviation functions} (LDFs) \cite{touchette09a} controlling the statistics of these fluctuations exhibit non-analyticities and Lee-Yang singularities \cite{yang52a,arndt00a,blythe02a,dammer02a,blythe03a,flindt13a,hickey14a,brandner17a} at the DPT reminiscent of standard critical behavior. This is a finding of crucial importance in nonequilibrium physics, as these LDFs play a role akin to the equilibrium thermodynamic potentials for nonequilibrium systems, where no bottom-up approach exists yet connecting microscopic dynamics with macroscopic properties \cite{bertini15a,derrida07a,barre17a}. Moreover, the emergence of coherent structures associated to rare fluctuations implies in turn that these extreme events are far more probable than previously anticipated \cite{lam09a,hurtado14a}. 

Despite their conceptual importance, observing DPTs is challenging as the spontaneous emergence of large fluctuations in macroscopic systems is unlikely \cite{bertini15a}, so one may question their physical relevance. However, recent breakthroughs have shown that fluctuations admit a control-theory (or active) interpretation \cite{bertini15a,chetrite15a,chetrite15b} where rare trajectories become typical under the action of an external control field. Among the fields that drive the system to the desired fluctuation, the one minimizing the dissipated energy is univocally related to the typical trajectory for the spontaneous emergence of such fluctuation \cite{bertini15a}. In this way, a DPT at the trajectory level corresponds to a singular change in the optimal control field, and this could be easily observed in actual experiments. In this sense DPTs are not only of conceptual but also of practical importance, specially for realistic $d>1$ systems \cite{pinchaipat17a,abou17a} amenable to control for technological applications. However, up to now most works on DPTs have focused on toy $1d$ models \cite{hurtado11a,perez-espigares13a,hurtado14a,vaikuntanathan14a,jack15a,shpielberg16a,zarfaty16a,baek17a,karevski17a,garrahan10a,ates12a,lesanovsky13a,carollo17a} or fluctuations of \emph{scalar} ($1d$) observables in $d>1$ \cite{garrahan07a,garrahan09a,hedges09a,chandler10a,pitard11a,speck12a,pinchaipat17a,abou17a,garrahan11a,genway12a,manzano14a}, and the challenge remains to understand DPTs in the fluctuations of fully vectorial observables in $d$-dimensions and how they are affected by the (possible) system anisotropy. 

In this paper we address this challenge and report compelling evidences of a rich DPT and new physics in the statistics of vectorial currents in an archetypal $2d$ driven diffusive system, the weakly asymmetric simple exclusion process (WASEP) \cite{derrida98a}. To crack this problem, we use massive cloning Monte Carlo simulations for rare event statistics \cite{giardina06a,lecomte07a,giardina11a}, together with macroscopic fluctuation theory (MFT) to understand the fluctuation phase diagram \cite{bertini15a}. We find a $2^{\text{nd}}$-order DPT between a homogeneous fluctuation phase with structureless trajectories and Gaussian current statistics, and a non-Gaussian phase for small currents. This non-Gaussian phase is characterized by the emergence of coherent jammed states in the form of traveling-wave trajectories, thus breaking the spatio-temporal translation symmetry. Such jammed states, which are surprisingly extended and non-compact, hamper particle flow enhancing the probability of low-current fluctuations \cite{perez-espigares13a}, and we introduce a novel order parameter for their detection. Interestingly, for mild or no anisotropy different symmetry-broken phases appear (depending on the current vector) separated by lines of $1^{\text{st}}$-order DPTs, a degeneracy which disappears beyond a critical anisotropy. Dynamical coexistence of the different traveling-wave phases appears along these $1^{\text{st}}$-order lines. 

{\bf \emph{Model}}-- 
The $2d$-WASEP belongs to a broad family of driven diffusive systems of fundamental and technological interest  \cite{bertini15a,derrida07a,hurtado14a}. Microscopically, this model is defined on a $2d$ square lattice of size $N=L\times L$ with periodic boundaries where $M\leq N$ particles evolve, so the global density is $\rho_0 = M/N$. Each lattice site may contain at most one particle, which performs stochastic jumps to neighboring empty sites along the $\pm{\alpha}$-direction ($\alpha=x,y$) at a rate $r^\alpha_{\pm}\equiv\text{exp}[\pm E_\alpha/L]/2$, with $\vE=(E_x,E_y)$ being an external field. For large $\vE$ and moderate system sizes, the field per unit length $\vE/L$ is strong enough to induce an \emph{effective anisotropy} in the medium \footnote{The rates $r^\alpha _{\pm }\equiv \text{exp}[\pm E_\alpha /L]/2$ converge for large $L$ to the standard ones found in literature \cite {bodineau05a,derrida07a}, namely $\frac{1}{2}(1 \pm E_\alpha /L)$, but avoid problems with negative rates for small $L$. Indeed, the hydrodynamic description of both variants of the model is identical in the thermodynamic limit. However, for finite, moderate values of $L$ the field per unit length ($\vE/L$) is too strong, leading to an \emph{effective anisotropy} in the system. In fact, by expanding the microscopic transition rate $r_\pm ^\alpha$ to second order in the field per unit length, i.e. $r_\pm ^\alpha \approx \frac{1}{2}[1\pm E_\alpha /L+\frac{1}{2}(E_\alpha /L)^2] + {\cal O}[(E_\alpha/L)^3]$, it is easy to show using a simple random walk argument that the second-order perturbation results in an effective increase of diffusivity and mobility along the field direction, and an associated decrease in the orthogonal direction.}, enhancing diffusivity and mobility along the field direction, an effect that can be accounted for in our theory below by an effective anisotropy parameter $\epsilon$. 

{\bf \emph{Trajectory statistics}}-- 
We are interested in the statistical physics of an ensemble of trajectories conditioned to a given total vector current $\vQ$ integrated over a long time $t$. In the spirit of equilibrium statistical mechanics, this trajectory ensemble is fully characterized by a \emph{dynamical partition function} $Z_t(\vlamb)=\sum_\vQ P_t(\vQ) \text{e}^{\vlamb\cdot\vQ}$, where $P_t(\vQ)$ is the probability of trajectories of duration $t$ with total current $\vQ$, or equivalently by the associated \emph{dynamical free energy} (dFE) $\mu(\vlamb)=\lim_{t\to \infty} t^{-1} \ln Z_t(\vlamb)$. The intensive vector $\vlamb$ is conjugated to the extensive current $\vQ$, in a way similar to the relation between temperature and energy in equilibrium systems. However, and unlike temperature, the parameter $\vlamb$ is non-physical and cannot be directly manipulated, a main difficulty when studying DPTs which can be however circumvented using the \emph{active interpretation} of fluctuation formulas \cite{bertini15a}. In any case, fixing $\vlamb$ is equivalent to conditioning the system to have an intensive current $\vq_\vlamb\equiv \vQ_\vlamb/t = \vnabla_\vlamb \mu(\vlamb)$, so by varying $\vlamb$ one can move from one dynamical phase to another. 

{\bf \emph{Macroscopic fluctuation theory}}-- 
At the mesoscopic level, driven diffusive systems like WASEP are characterized by a density field $\rho(\vrr,t)$ obeying a continuity equation $\partial_t \rho + \vnabla \cdot \vj =0$, with a current field $\vj (\vrr,t)\equiv -\hD(\rho) \vnabla \rho + \hS(\rho)\vE + \vxi$. The field $\vxi(\vrr,t)$ is a Gaussian white noise of \emph{weak} amplitude $\propto L^{-1}$ (the inverse system size) which accounts for microscopic random fluctuations at the mesoscopic level, and $\vE$ is the external field driving the system out of equilibrium. The deterministic part of $\vj (\vrr,t)$ is given by Fick's law, with $\hD(\rho)\equiv D(\rho)\mA$ and $\hS(\rho)=\sigma(\rho)\mA$ the diffusivity and mobility matrices, respectively. The constant diagonal matrix $\mA$ measures the system underlying anisotropy, i.e. the possible change of microscopic jump rates from one spatial direction to another. We are interested in the statistics of trajectories $\{\rho(\vrr,t),\vj(\vrr,t)\}_0^\tau$ constrained to a fixed current $\vq=\tau^{-1}\int_0^\tau dt \int d\vrr \, \vj(\vrr,t)$  during a long time $\tau$ in a closed system with periodic boundaries. The associated nonequilibrium steady state is homogeneous, with constant (and conserved) density $\rho_0$ and average current $\la\vq\ra=\sigma_0\mA\vE$, with $\sigma_0\equiv \sigma(\rho_0)$. MFT offers precise variational formulas for the dFE $\mu(\vlamb)$ starting from the above fluctuating hydrodynamics equations \cite{bertini15a}, and with the only input of two transport coefficients, which for $2d$-WASEP are $D(\rho)=1/2$ and $\sigma(\rho)=\rho(1-\rho)$, and an anisotropy matrix that we parametrize here as $\mA_{\text{xx}}=1+\epsilon$ and $\mA_{\text{yy}}=1-\epsilon$. This MFT problem can be solved using standard techniques, see Supplementary Material \cite{SMprl}, and we now summarize its predictions. 

\begin{figure}[t]
\vspace{-0.3cm}
\includegraphics[width=8.5cm]{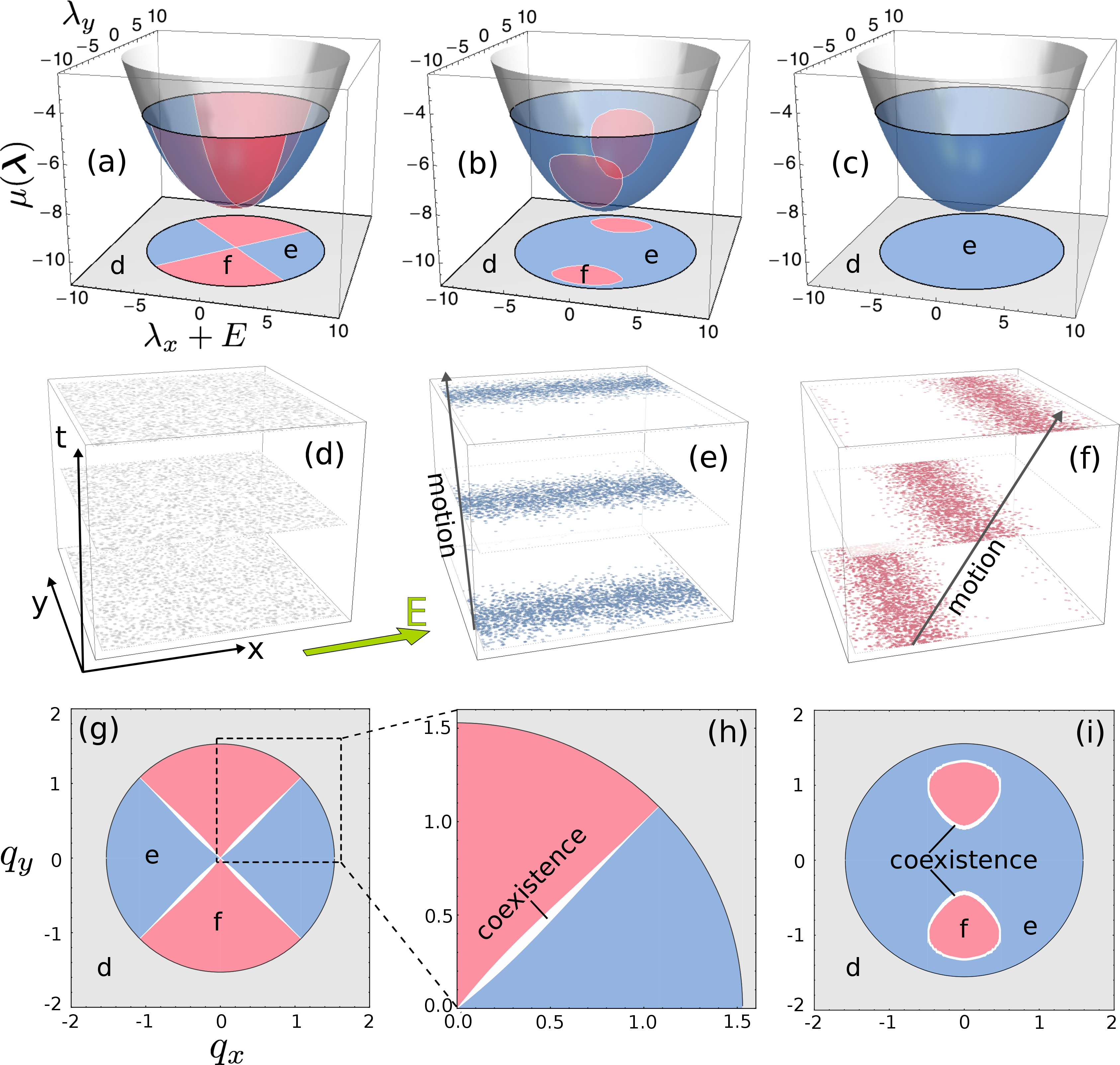}
\vspace{-0.2cm}
\caption{(Color online) Top row: $\mu(\vlamb)$ for the $2d$-WASEP in an external field $\vE=(10,0)$, as derived from MFT, in the case of (a) no anisotropy, $\epsilon=0$, (b) mild anisotropy, $0<\epsilon<\epsilon_c$, and (c) strong anisotropy,  $\epsilon>\epsilon_c$. The projections show the phase diagram in $\vlamb$-space for each case, and letters indicate the typical spatiotemporal trajectories in each phase, displayed in the middle row (d)-(f). A DPT appears between a Gaussian phase (light gray) with homogeneous trajectories (d) and two different non-Gaussian symmetry-broken phases for low currents characterized by jammed density waves, (e) and (f). The first DPT is $2^{\text{nd}}$-order, while the two symmetry-broken phases are separated by lines of $1^{\text{st}}$-order DPTs. Bottom row: phase diagram in current space for anisotropy $\epsilon=0$ (g,h), and $0<\epsilon<\epsilon_c$ (i). The coexistence pockets (white) are apparent. 
}
\label{fig1}
\end{figure}

{\bf \emph{Dynamical phase diagram}}-- Small current fluctuations ($|\vq-\la\vq\ra|\ll1$ or $|\vlamb|\approx 0$) typically result from the random superposition of mostly-independent local jumps which sum incoherently to yield the desired current, so the typical trajectories associated to these small fluctuations are still homogeneous, as the stationary ones \cite{bodineau05a,hurtado11a}. According to the central limit theorem, this leads to Gaussian current statistics corresponding to a quadratic dynamical free energy $\mu_{\text{G}}(\vz)\equiv (\vz\cdot \hS_0 \vz - \vE\cdot\hS_0\vE)/2$, with $\vz\equiv \vlamb+\vE$. This homogeneous phase is depicted in light gray in Fig.~\ref{fig1}. A local stability analysis then shows that this Gaussian, homogeneous regime eventually becomes unstable against small but otherwise arbitrary spatiotemporal perturbations in trajectories. For WASEP this happens for large enough external fields and currents $\vq\cdot\mA^{-1}\vq \le \sigma_0^2\Xi_c$, or equivalently $\vz\cdot\mA\vz\le \Xi_c$, where $\Xi_c$ is a critical threshold, see black lines separating gray and colored regions in Fig.~\ref{fig1}.a-c. This transition can be shown to be of $2^{\text{nd}}$-order type as $\partial^2_{|\vz|}\mu(\vz)$ is discontinuous at the critical line \cite{SMprl}. 

Interestingly, the dominant perturbation immediately after the instability kicks in takes the form of a traveling density wave with structure only along one-dimension ($1d$), either $x$ or $y$  (see Figs.~\ref{fig1}.e-f). This collective rearrangement breaks the system spatiotemporal translation symmetry by localizing particles in a jammed region to facilitate a low-current fluctuation. This solution can be extended to all currents below the critical line, and we find that different $1d$ density waves dominate different current vector regimes, depending on the anisotropy parameter $\epsilon$, see Figs.~\ref{fig1}.a-c. Lines of $1^{\text{st}}$-order DPTs separate both density wave phases where the dFE $\mu(\vlamb)$ exhibits a jump in its first derivative \cite{SMprl}, so the current $\vq_\vlamb = \vnabla_\vlamb \mu(\vlamb)$ corresponding to a given $\vlamb$ jumps discontinuously at these lines. In this way the $1^{\text{st}}$-order DPT lines in $\vlamb$-space correspond to pockets in $\vq$-space where \emph{dynamical coexistence} emerges between the two traveling-wave phases, see Fig. \ref{fig1}.g-i. This means that if we were to observe an atypical current $\vq$ sitting in one of these pockets, either by an unlikely spontaneous fluctuation or by an active control of the current with an optimal field, we would observe dynamical coexistence of the two different traveling density waves.

Strikingly, particular $2d$ traveling-wave solutions (as e.g. traveling \emph{compact} packets) do not improve the variational problem for $\mu(\vlamb)$ when compared to their $1d$ counterparts. This is surprising, as one would naively expect the system to minimize the interface between the high- and low-density regions while developing a macroscopic jam to sustain a low-current fluctuation. This phenomenological picture does not emerge in our theory and is not observed in simulations below.

What are the key ingredients responsible of the new physics here described and not observed in previous works \cite{hurtado11a,perez-espigares13a,hurtado14a,vaikuntanathan14a,jack15a,shpielberg16a,zarfaty16a,baek17a,karevski17a,garrahan10a,ates12a,lesanovsky13a,carollo17a,garrahan07a,garrahan09a,hedges09a,chandler10a,pitard11a,speck12a,pinchaipat17a,abou17a,garrahan11a,genway12a,manzano14a}? First, by considering \emph{vectorial currents} it becomes apparent that current rotations can trigger $1^{\text{st}}$-order transitions between different symmetry-broken jammed dynamical phases. This is certainly not present in simpler $1d$ models \cite{hurtado11a,perez-espigares13a,hurtado14a,vaikuntanathan14a,jack15a,shpielberg16a,zarfaty16a,baek17a,karevski17a,garrahan10a,ates12a,lesanovsky13a,carollo17a} and cannot show up when studying fluctuations of scalar observables in $d>1$ \cite{garrahan07a,garrahan09a,hedges09a,chandler10a,pitard11a,speck12a,pinchaipat17a,abou17a,garrahan11a,genway12a,manzano14a}. Second, by including \emph{anisotropy} in our analysis (a main feature of many realistic $d>1$ systems not considered before), it becomes clear its strong effect on the relative shape and position of the different jammed phases, see Fig. \ref{fig1}.a-c. In this way, it is the interplay between vectorial currents and anisotropy in $d>1$ what gives rise to the rich and complex dynamical phase diagram here described. Mathematically, the novel competition between different symmetry-broken dynamical phases is due to the appearance of a structured vector field coupled to the current \cite{perez-espigares16a,villavicencio16a,tizon-escamilla17a}.

\begin{figure}[t]
\vspace{-0.3cm}
\centerline{\includegraphics[width=9cm]{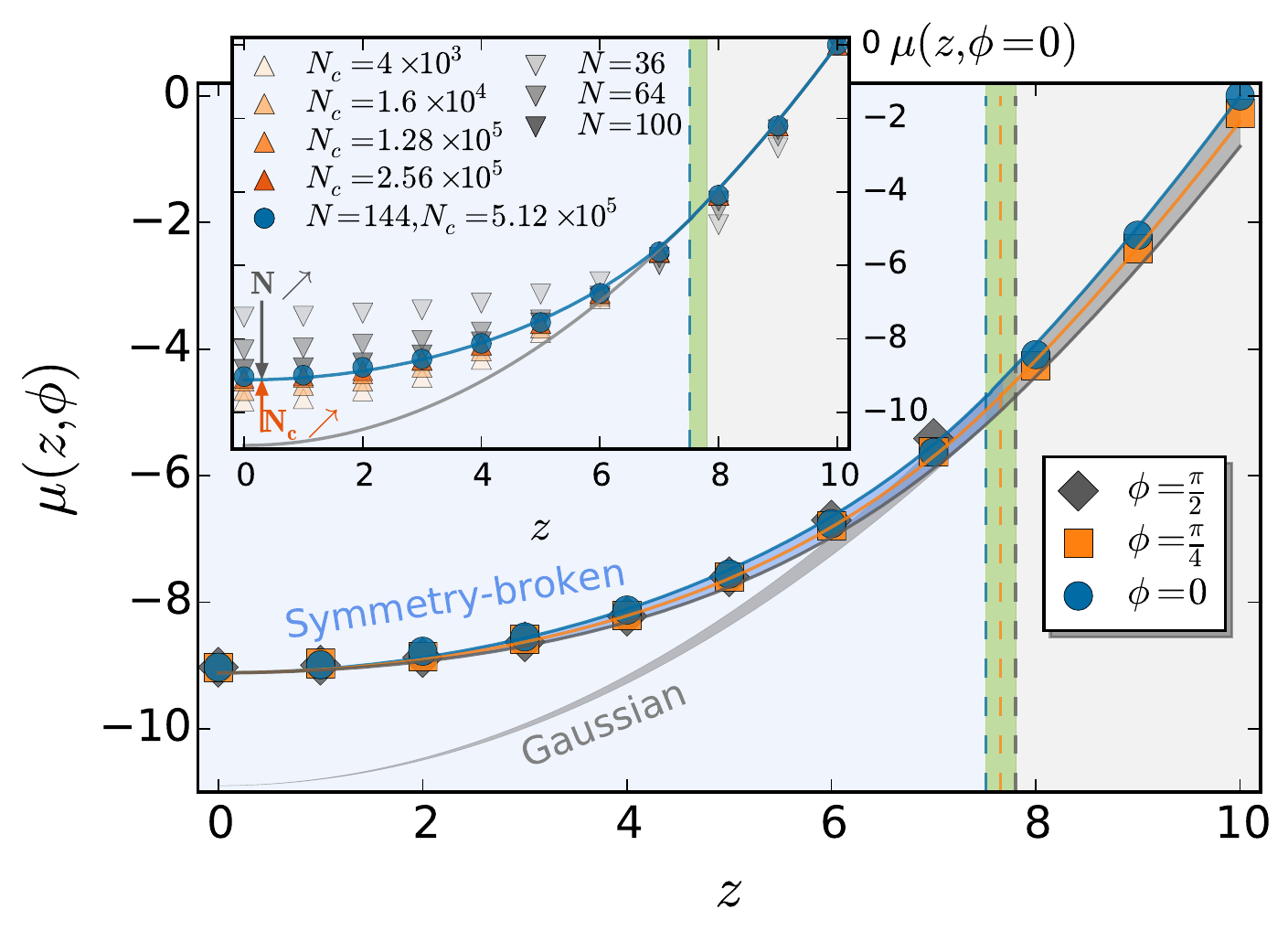}}
\vspace{-0.4cm}
\caption{(Color online) Main: $\mu(\vlamb)$ vs $z=|\vlamb+\vE|$ as obtained in simulations for $N=144$, $N_c=5.12\times10^5$ and different $\phi=\tan^{-1}(z_y/z_x)$, together with MFT predictions for anisotropy $\epsilon=0.038$. A DPT from a Gaussian regime (light-gray ribbon) to a symmetry-broken, non-Gaussian phase (blue ribbon) is apparent upon crossing $z_c(\phi)$, with $\vz_c\cdot\mA\vz_c=\Xi_c$ (green vertical stripe). Different $\phi$ correspond to different MFT lines within the shaded ribbons. Inset: Convergence to the $\phi=0$ MFT prediction (blue line) for $N=144$ as $N_c$ increases ($\bigtriangleup$) and for optimal $N_c$ as $N$ increases  ($\bigtriangledown$). 
}
\label{fig2}
\end{figure}

{\bf \emph{Numerical results}}-- The previous results call for independent numerical verification, as they derive from an effective mesoscopic theory which relies on a few hypotheses \cite{bertini15a,SMprl}. To search for this DPT, we explored the current statistics of the $2d$-WASEP using massive cloning Monte Carlo simulations \cite{giardina06a,lecomte07a,giardina11a}. In particular, we simulated systems with density $\rho_0=0.3$, several system sizes up to $N=144$, and a strong external field $\vE=(10,0)$. The cloning Monte Carlo method relies on a controlled modification of the system stochastic dynamics such that the rare events responsible for a given fluctuation are no longer rare, and involves the parallel simulation of multiple copies of the system \cite{giardina06a,lecomte07a,giardina11a}. The number of clones needed to observe a given rare event grows exponentially with the system size, all the more the rarer the event is \cite{hurtado09a,nemoto16a}. In particular, to pick up and characterize reliably the DPT in the $2d$-WASEP we needed the extraordinary number of $N_c=5.12\times10^5$ clones evolving in parallel for a long time. 

\begin{figure}[t]
\vspace{-0.3cm}
\centerline{\includegraphics[width=8.5cm]{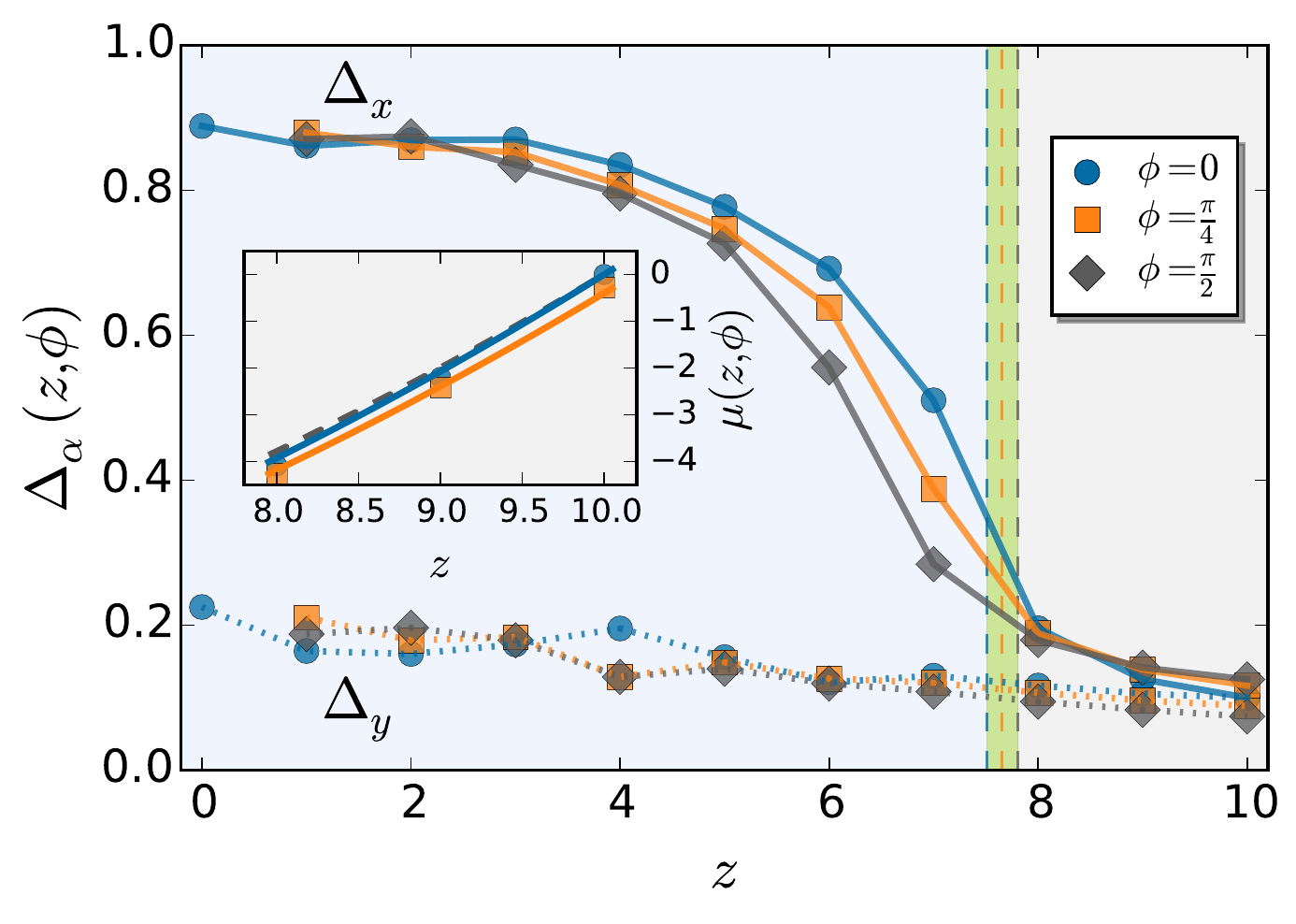}}
\vspace{-0.5cm}
\caption{(Color online) Tomographic $\alpha$-coherences, with $\alpha=x,y$, as a function of $z$ for different current angles $\phi$ measured for $N=100$ and $\vE=(10,0)$. Inset: dFE $\mu(\vz)$ vs $z$ in the Gaussian regime for $\phi=0,\pi/4$, see Fig.~\ref{fig2}. Full (dashed) lines are MFT predictions with anisotropy $\epsilon=0.038$ ($\epsilon=0$).}
\label{fig3}
\end{figure}

According to MFT, Gaussian current statistics corresponding to a quadratic dFE $\mu_{\text{G}}(\vz)$ are expected for $\vz\cdot\mA\vz\ge\Xi_c$, see Fig.~\ref{fig1} and discussion above. This is fully confirmed in Fig.~\ref{fig2}, which shows the measured $\mu(\vz)$ for $N=144$ as a function of $z=|\vz|$ for different current orientations $\phi=\tan^{-1}(z_y/z_x)$. This confirms that mild current fluctuations stem from the random superposition of weakly-correlated, localized events which sum up incoherently to yield Gaussian statistics.  Interestingly, we find a weak dependence of $\mu(\vz)$ on $\phi$ in this Gaussian regime, a clear hallmark of the effective anisotropy mentioned above. Indeed, this $\phi$-dependence can be used to estimate that $\epsilon\approx 0.038$ properly describes the observed weak anisotropy, see inset in Fig.~\ref{fig3}. This effective anisotropy is slightly larger than the critical anisotropy $\epsilon_c\approx 0.035$ beyond which a single symmetry-broken phase dominates the non-Gaussian regime, see Fig.~\ref{fig1}.c, an observation consistent with additional results below. The Gaussian, incoherent fluctuation regime ends up for $\vz\cdot\mA\vz<\Xi_c$, where clear deviations from the quadratic form $\mu_{\text{G}}(\vz)$ become apparent, see Fig.~\ref{fig2}. This change of behavior, in excellent agreement with MFT predictions, signals the onset of the DPT to a symmetry-broken phase characterized by non-Gaussian current fluctuations and traveling density wave trajectories. A clear convergence to the MFT prediction is observed in the Gaussian and non-Gaussian regimes as both $N$ and the number of clones $N_c$ increase, see inset in Fig.~\ref{fig2}. 

The smoking gun of any continuous phase transition, such as the DPT here reported, is a smooth but apparent change in an order parameter \cite{binney92a}. To distinguish between the different jammed density-wave phases which are expected to appear for low current fluctuations, see Fig.~\ref{fig1}.e-f, we introduce now a \emph{structural} order parameter capable of discerning the jam direction, if any (see \cite{SMprl} for a detailed description). In particular, we take $1d$ \emph{slices} of our $2d$ system along a given direction, $\alpha=x$ or $y$, and compute the center of mass position for each slice. Clearly, a small average dispersion $\la\sigma_\alpha^2\ra_\vlamb$ of the centers of mass across the different slices signals the formation of a jam along the $\alpha$-direction, Fig.~\ref{fig1}.e-f, while random homogeneous configurations typical of the Gaussian phase (Fig.~\ref{fig1}.d) are characterized by a large dispersion. We hence define the \emph{tomographic $\alpha$-coherence} (i.e. the center-of-mass coherence across the different slices along the $\alpha$-axis) as $\Delta_\alpha(\vlamb)\equiv 1-\la\sigma_\alpha^2\ra_\vlamb$, and Fig.~\ref{fig3} shows this order parameter measured in simulations across the DPT for $\alpha=x,y$. Remarkably, $\Delta_x(z)$ increases steeply for $\vz\cdot\mA\vz\le \Xi_c$ and \emph{all angles} $\phi$ of the current vector, while $\Delta_y(z)$ remains small and does not change appreciably across the DPT, clearly indicating that only one of the two possible symmetry-broken phases appear in our simulations, as expected from MFT in the supercritical anisotropy regime $\epsilon>\epsilon_c$ and consistent with the measured effective anisotropy $\epsilon\approx0.038>\epsilon_c$, see inset in Fig.~\ref{fig3}. Note also that the behavior of both $\Delta_\alpha(z)$ across the DPT is consistent with the emergence of a traveling wave with structure in $1d$ and not in $2d$, as in the latter case both $\Delta_\alpha(z)$ should increase upon crossing $z_c(\phi)$. Moreover, the steep but continuous change of $\Delta_x(\vz)$ across the DPT is consistent with a second-order transition, in agreement with MFT.

{\bf \emph{Summary}}-- We have presented compelling evidences of a complex dynamical phase transition in the current vector statistics of a paradigmatic model of transport in $2d$, characterizing its properties with the tools of macroscopic fluctuation theory. Our analysis of MFT equations predicts a rich phase diagram, with non-analiticities of $1^{\text{st}}$- and $2^{\text{nd}}$-order type in the current dynamical free energy, accompanied by emergent order in different symmetry-broken phases characterized by traveling density waves. This richness is aided by the complex interplay among anisotropy, external field and vector currents in $d>1$, key features missing in the simpler models studied in the past. Interestingly, our results show that order and coherence may emerge out of an unlikely fluctuation, proving the deep connection between rare events and self-organized structures which enhance their probability. This is expected to be a general feature of many complex dynamical systems \cite{lam09a}. The mapping between exclusion processes and dual quantum spin systems \cite{felderhof71a,alcaraz94a,stinchcombe95a,stinchcombe01a} suggests a connection between the DPT here uncovered and a rich quantum phase transition yet to be explored. It would be also interesting to determine the universality class of this DPT, and the dynamical exponents of the different fluctuation phases \cite{karevski17a,lecomte07b}.

\begin{acknowledgments}
Financial support from Spanish projects FIS2013-43201-P (MINECO) and FPU13/05633, Italian Research Funding Agency (MIUR) through FIRB project grant RBFR10N90W, Italian INdAM \emph{Francesco Severi}, University of Granada, Junta de Andaluc\'{\i}a project P09-FQM4682 and GENIL PYR-2014-13 project is acknowledged.
\end{acknowledgments}


%

\newpage

\appendix

\onecolumngrid


\section{Dynamic phase transitions in the current vector statistics from macroscopic fluctuation theory}

In this section we analyze the equations of macroscopic fluctuation theory (MFT) for the current vector statistics of arbitrary driven diffusive systems, with special emphasis on the MFT predictions regarding the existence and nature of dynamic phase transitions (DPTs) in some regimes of current fluctuations. In particular, we consider a broad class of $d$-dimensional anisotropic driven diffusive systems characterized by a locally-conserved density field $\rho(\vrr,t)$ which evolves in time according to the following fluctuating hydrodynamics equation \cite{Bertini1,Derrida1,weJSP}
\be
\partial_t \rho(\vrr,t) + \vnabla \cdot \left( -\hD(\rho) \vnabla \rho(\vrr,t) + \hS(\rho)\vE + \vxi(\vrr,t) \right) = 0 \, ,
\label{langevin}
\ee
with $\vE$ the external field driving the system out of equilibrium and $\vrr\in \Lambda\equiv [0,1]^d$. The field $\vj (\vrr,t)\equiv -\hD(\rho) \vnabla \rho(\vrr,t) + \hS(\rho)\vE + \vxi(\vrr,t)$ is the fluctuating current, with $\hD(\rho)\equiv D(\rho)\mA$ and $\hS(\rho)=\sigma(\rho)\mA$ the diffusivity and mobility matrices, respectively, and $\mA$ a diagonal anisotropy matrix with components $\mA_{\alpha\beta}=a_\alpha\delta_{\alpha\beta}$, $\alpha,\beta\in [1,d]$. The noise term $\vxi(\vrr,t)$ is Gaussian and white with zero average, $\la \vxi(\vrr,t)\ra=0$, and variance 
\be
\la \xi_\alpha(\vrr,t)\xi_\beta(\vrr',t') \ra=L^{-d}\sigma(\rho) a_\alpha\delta_{\alpha\beta}\delta(\vrr-\vrr')\delta(t-t') \, , \nonumber
\ee
with $L$ the system size in natural units. This (conserved) noise term accounts for the many fast microscopic degrees of freedom which are averaged out in the coarse-graining procedure resulting in Eq. (\ref{langevin}). The diffusion and mobility transport matrices fully characterize the macroscopic fluctuation properties of the model at hand, being related via a local Einstein relation $\hD(\rho)=f_0''(\rho) \hS(\rho)$, with $f_0(\rho)$ the \emph{equilibrium} free energy of the system. To completely define the problem, the evolution equation (\ref{langevin}) must be supplemented with appropriate boundary conditions, which in this case are simply periodic along all $d$ directions.

Now, starting from the Fokker-Planck description of the Langevin equation (\ref{langevin}) and using a path integral formalism, the probability of observing a given trajectory $\{\rho(\vrr,t),\vj(\vrr,t)\}_0^{\tau}$ of duration $\tau$ for the density and current fields can be written as \cite{Bertini1}
\be
\text{P}\left(\{\rho,\vj\}_0^{\tau} \right) \asymp \exp \Big( +L^d I_{\tau}\left[\rho,\vj \right] \Big) \, , 
\label{probpath} 
\ee
where the symbol "$\asymp$" stands for asymptotic logarithmic equality, i.e.
\be
\lim_{L\to\infty} \frac{1}{L^d} \ln \text{P}\left(\{\rho,\vj\}_0^{\tau} \right) = I_{\tau}\left[\rho,\vj \right] \, .
\label{asymp}
\ee
The action of Eq. (\ref{probpath}) is
\be
I_{\tau}\left[\rho,\vj \right]=- \int_0^{\tau} dt \int_\Lambda d \vrr\, \frac{1}{2\sigma(\rho)}\left(\vj+D(\rho)\mA \vnabla\rho -\sigma(\rho)\mA\vE \right) \cdot \mA^{-1} \left(\vj+D(\rho)\mA \vnabla\rho -\sigma(\rho)\mA\vE \right) \,  , \nonumber
\ee
where the fields $\rho (\vrr,t)$ and $\vj(\vrr,t)$ are coupled via the continuity equation, see Eq. (\ref{langevin}),
\be
\partial_t \rho (\vrr,t) + \vnabla\cdot \vj(\vrr,t) = 0 \, . 
\label{contappA}
\ee
For any other trajectory not obeying (\ref{contappA}), $I_{\tau}\left[\rho,\vj \right]\to-\infty$. Moreover, the system of interest is isolated so that the total mass is conserved,
\be
\rho_0=\int_{\Lambda}d\vrr\, \rho(\vrr,t) \, . 
\label{massappA}
\ee
The probability $\text{P}_{\tau}(\vq)$ of observing a space- and time-averaged empirical current \emph{vector} $\vq$, defined as
\be
\vq = \frac{1}{\tau}  \int_0^{\tau} dt \int_\Lambda d\vrr \, \vj (\vrr,t) \, ,
\label{currappA}
\ee
scales for long times as $\text{P}_{\tau}(\vq)\asymp \exp[+\tau L^d G(\vq)]$, and the current large deviation function (LDF) $G(\vq)$ can be related to $I_{\tau}[\rho,\vj]$ via a simple saddle-point calculation in the long-time limit, 
\be
G(\vq) = \lim_{\tau \to \infty} \frac{1}{\tau} \max_{\{\rho,\vj\}_0^\tau} I_{\tau}[\rho,\vj] \, ,
\label{LDFappA}
\ee
subject to constraints (\ref{contappA}), (\ref{massappA}) and (\ref{currappA}). The density and current fields solution of this variational problem, denoted here as $\rho_\vq(\vrr,t)$ and $\vj_\vq(\vrr,t)$, correspond to the optimal path the system follows in mesoscopic phase space to sustain a long-time current fluctuation $\vq$. This path may be in general time-dependent, and the associated general variational problem is remarkably hard. 

This problem becomes simpler however in different limiting cases. For instance, in the steady state the system exhibits translation symmetry with an homogeneous stationary density profile $\rho_{\text{st}}(\vrr)=\rho_0$ and a constant average current $\vj_{\text{st}}(\vrr)=\la\vq\ra=\sigma_0\mA \vE$, where we have defined $\sigma_0\equiv \sigma(\rho_0)$. Now, one can argue that small fluctuations of the empirical current $\vq$ away from the average behavior $\la \vq\ra$ will typically result from weakly-correlated local events in different parts of the system which add up incoherently to yield the desired $\vq$, so the optimal density field associated to these small fluctuations still corresponds to the homogeneous, stationary one \cite{BD2,weSSB}, i.e. $\rho_\vq(\vrr,t)=\rho_0$ for $|\vq-\la\vq\ra|\ll1$, while the optimal current field is constant, $\vj_\vq(\vrr,t)=\vq$, leading to a quadratic current LDF corresponding to Gaussian current statistics,
\beq
G_{\text{G}}(\vq)=-\frac{1}{2\sigma_0}\left(\vq-\sigma_0\mA\vE\right)\cdot\mA^{-1}\left(\vq-\sigma_0\mA\vE\right) \, ,
\label{Gqgaus}
\eeq
as indeed corroborated in our simulations for a broad range of $\vq$'s. As an interesting by-product, note that current fluctuations in this Gaussian regime obey an anisotropic version of the Isometric Fluctuation Theorem \cite{weIFR,AFR,SFT}, which links in simple terms the probability of two different but $\mA$-isometric current vector fluctuations. In particular, 
\be
\lim_{\tau\to\infty}\frac{1}{\tau L^d} \ln\left[ \frac{P_\tau(\vq)}{P_\tau(\vq')}\right]= \vE\cdot(\vq-\vq') \, , 
\label{IFRgauss}
\ee
$\forall \vq,\vq'$ in the Gaussian regime such that $\vq\cdot\mA\vq = \vq'\cdot\mA\vq'$.

Interestingly, the above ansatz with the associated \emph{flat} profiles remains a solution of the full variational problem $\forall \vq$, but the question remains as to whether other solutions with more complex spatiotemporal structure may yield a better maximizer of the MFT action (\ref{LDFappA}) for currents. To address this question, we now perturb the above flat solution with small but otherwise arbitrary functions of space and time, and study the local stability of the homogeneous solution against such perturbations. In particular, we ask whether the perturbed fields yield in some case a larger $G(\vq)$. With this aim in mind, we write
\be
\bar{\rho}(\vrr,t)=\rho_0+\delta\rho(\vrr,t), \ \ \ \ \bar{\vj}(\vrr,t)=\vq+\delta\vj(\vrr,t) \, ,
\ee
where both $\bar{\rho}(\vrr,t)$ and $\bar{\vj}(\vrr,t)$ remain constrained by Eqs. (\ref{contappA}), (\ref{massappA}) and (\ref{currappA}). Inserting these expressions in Eq. (\ref{LDFappA}) and expanding to second order in the perturbations, we obtain the leading correction to the quadratic form $G_{\text{G}}(\vq)$ of Eq. (\ref{Gqgaus}) (termed here $O2$)
\be
\displaystyle O2=-\frac{1}{2\tau}\int_0^{\tau}dt\int_{\Lambda}d\vrr\left\lbrace A(\rho_0,\vq) \delta\rho^2+
\displaystyle \vnabla\delta\rho \cdot\hat{B}(\rho_0)\vnabla\delta\rho+
\displaystyle \delta\vj \cdot\hat{C}(\rho_0)\delta\vj +\delta\vj \cdot{\bf{F}}(\rho_0,\vq)\delta\rho \right\rbrace \, ,
\label{O2flat}
\ee
where we have defined
\be
A(\rho_0,\vq)=\left( \frac{\sigma_0'^2}{\sigma_0^3} - \frac{\sigma''_0}{2\sigma_0^2} \right) \vq\cdot\mA^{-1}\vq + \sigma''_0\vE\cdot\mA\vE ,\quad \hat{B}(\rho_0)=\frac{D_0^2}{\sigma_0}\mA ,\quad \hat{C}(\rho_0)=\frac{\mA^{-1}}{\sigma_0} ,\quad {\bf{F}}(\rho_0,\vq)=-\frac{\sigma_0'}{\sigma_0^2}\mA^{-1}\vq \, ,
\ee
with $'$ denoting derivative with respect to the argument, and $D_0\equiv D(\rho_0)$. We next expand the perturbations $\delta\rho(\vrr,t)$ and $\delta\vj(\vrr,t)$ in Fourier series, taking advantage of the spatial periodic boundary conditions, and imposing explicitly along the way the constraints (\ref{contappA}), (\ref{massappA}) and (\ref{currappA}). For simplicity we particularize hereafter our results for dimension two, $d=2$, though the generalization to arbitrary $d$ is straightforward. In this way, perturbations take the form
\begin{eqnarray}
\displaystyle\delta\rho(\vrr,t)&=&\sum_\nu\frac{1}{\nu}\Big[-\nabla\cdot\bm{\gamma}_{1,\nu}(\vrr)\sin(\nu t)+\nabla\cdot\bm{\gamma}_{2,\nu}(\vrr)\cos(\nu t)\Big] \, ,\\
\displaystyle\delta\vj(\vrr,t)&=&\sum_\nu\Big[\bm{\gamma}_{1,\nu}(\vrr)\cos(\nu t)+\bm{\gamma}_{2,\nu}(\vrr)\sin(\nu t)\Big] \, ,
\end{eqnarray}
where the first equation follows from the second expansion after imposing the continuity constraint (\ref{contappA}), with
\begin{eqnarray}
&\displaystyle\bm{\gamma}_{1,\nu}(\vrr)=\frac{1}{4}{\bf a}_{\nu00}+\frac{1}{2}\sum_{k_1\ne 0}\left({\bf a}_{\nu k_10}\cos{k_1x}+{\bf c}_{\nu k_10}\sin{k_1x}\right)+\frac{1}{2}\sum_{k_2\ne 0}\left({\bf a}_{\nu0k_2}\cos{k_2y}+{\bf b}_{\nu0k_2}\sin{k_2y}\right) + \\\nonumber
&+ \displaystyle\sum_{k_1,k_2\ne 0}\left({\bf a}_{\nu k_1k_2}\cos{k_1x}\cos{k_2y}+{\bf b}_{\nu k_1k_2}\cos{k_1x}\sin{k_2y}+{\bf c}_{\nu k_1k_2}\sin{k_1x}\cos{k_2y}+{\bf d}_{\nu k_1k_2}\sin{k_1x}\sin{k_2y}\right)
\end{eqnarray}
\begin{eqnarray}
&\displaystyle\bm{\gamma}_{2,\nu}(\vrr)=\frac{1}{4}{\bf s}_{\nu00}+\frac{1}{2}\sum_{k_1\ne 0}\left({\bf s}_{\nu k_10}\cos{k_1x}+{\bf u}_{\nu k_10}\sin{k_1x}\right)+\frac{1}{2}\sum_{k_2\ne 0}\left({\bf s}_{\nu0k_2}\cos{k_2y}+{\bf t}_{\nu0k_2}\sin{k_2y}\right) + \\\nonumber
&+ \displaystyle\sum_{k_1,k_2\ne 0}\left({\bf s}_{\nu k_1k_2}\cos{k_1x}\cos{k_2y}+{\bf t}_{\nu k_1k_2}\cos{k_1x}\sin{k_2y}+{\bf u}_{\nu k_1k_2}\sin{k_1x}\cos{k_2y}+{\bf v}_{\nu k_1k_2}\sin{k_1x}\sin{k_2y}\right)
\end{eqnarray}
where ${\bf a}_{\nu ij}$, ${\bf b}_{\nu ij}$, ${\bf c}_{\nu ij}$, ${\bf d}_{\nu ij}$, ${\bf s}_{\nu ij}$, ${\bf t}_{\nu ij}$, ${\bf u}_{\nu ij}$, ${\bf v}_{\nu ij}$ are the coefficients of the Fourier series. Note that the previous expansion has been divided into first the only-temporal modes, then all $1+1$ spatiotemporal modes along each direction of space, and finally the fully $2+1$ spatiotemporal modes. The $O2$ correction (\ref{O2flat}) is of course a quadratic form of the perturbations with constant coefficients, so the different Fourier modes decouple simplifying the problem. In this way the stability analysis melts down as usual to an eigenvalue problem, which in this case splits into different problems for only temporal modes, spatiotemporal modes with structure along just one dimension, $x$ or $y$, and $2d$ spatiotemporal modes, which can be analyzed separately. This straightforward but lengthy calculation leads to the following conclusion: the flat solution corresponding to Gaussian current statistics remains stable (i.e. the $O2$ correction is negative) whenever the following conditions hold,
\begin{eqnarray}\nonumber
\displaystyle a_{\text{min}}k_n^2\frac{D_0^2}{\sigma_0}&+&H(\vE,\vq)>0\\
\displaystyle a_{\text{max}}k_m^2\frac{D_0^2}{\sigma_0}&+&H(\vE,\vq)>0 \label{staresul}\\\nonumber
\displaystyle\left(a_{\text{min}}k_n^2+a_{\text{max}}k_m^2\right)\frac{D_0^2}{\sigma_0}&+&H(\vE,\vq)>0,
\end{eqnarray}
with $k_n=2\pi n$ and $k_m=2\pi m$ the different spatial modes associated to each perturbation along either direction, $a_{\text{min}}=\min\lbrace a_{\alpha}, \alpha\in[1,d]\rbrace$ and $a_{\text{max}}=\max\lbrace a_{\alpha}, \alpha\in[1,d]\rbrace$, and
\beq
H(\vE,\vq)=\frac{\sigma''_0}{2}\left(\vE\cdot\mA\vE-\sigma_0^{-2}\vq\cdot\mA^{-1}\vq\right)
\eeq
A number of important conclusions can be directly derived from this set of conditions, namely:
\begin{itemize}
\item[(i)] The first mode to become unstable (if any) is always the fundamental mode $k_1=2\pi$.
\item[(ii)] For any value of the anisotropy, the first perturbations to become unstable are those with structure along one spatial dimension, $x$ or $y$.
\item[(iii)] For anisotropic systems, $a_{\text{min}}<a_{\text{max}}$, the leading unstable perturbation has structure in the direction of minimum anisotropy. 
\item[(iv)] For isotropic systems, $a_{\text{min}}=a_{\text{max}}\equiv a$, both one-dimensional perturbations trigger the instability of the flat solution at the same point. In this case, the orientation of the current vector $\vq$ determines the most probable profile immediately after the instability kicks in, with structure only along the $x$- or $y$-direction, as dictated by the term proportional to ${\bf{F}}(\rho_0,\vq)$ in the $O2$ correction, see Eq. (\ref{O2flat}).
\end{itemize}
Therefore there exists a line of critical values for the current $\vq_c$ at which the instability appears, given by
\beq
\vq_c\cdot\mA^{-1}\vq_c=\sigma_0^2\left(\vE\cdot\mA\vE + 8\pi^2 a_{\text{min}}\frac{D_0^2}{\sigma_0\sigma_0''}\right)\equiv \sigma_0^2\Xi_c \, .
\label{xic}
\eeq
For systems with $\sigma_0''>0$ (as e.g. the Kipnis-Marchioro-Presutti model of heat transport \cite{weJSP,weIFR,kmp}), the instability appears always, regardless of the value of the external field (even for $\vE=0$), separating a regime of Gaussian current statistics for $\vq\cdot\mA^{-1}\vq \le \sigma_0^2\Xi_c$ and a non-Gaussian region for $\vq\cdot\mA^{-1}\vq > \sigma_0^2\Xi_c$. On the other hand, for systems with $\sigma_0''<0$ (as the weakly asymmetric simple exclusion process --WASEP-- studied in this paper \cite{wasep,BD2,wePRE}) a line of critical values of the external field exists, defined by
\beq
\vE_c\cdot\mA\vE_c=8\pi^2 a_{\text{min}}\frac{D_0^2}{\sigma_0|\sigma_0''|} \equiv |\Sigma_c|\, .
\label{ec}
\eeq
beyond which the instability appears, $\vE\cdot\mA\vE \ge |\Sigma_c|$. In this strong field case, Gaussian statistics are expected for all currents except for a region around $\vq=0$, defined by $\vq\cdot\mA^{-1}\vq \le \sigma_0^2\Xi_c$, where current fluctuations are non-Gaussian. For weak external fields, $\vE\cdot\mA\vE < |\Sigma_c|$, only Gaussian statistics are observed. 

Whenever the instability emerges, the first two frequencies to become unstable are $\nu_c^\pm=\pm 2\pi q_{\parallel}\sigma'_0/\sigma_0$, with $q_{\parallel}$ the component of the current vector along the direction of structure formation (that we denote here as $x_\parallel$). Considering that the first unstable spatial mode correspond to $k_\perp=0$, $k_\parallel=2\pi$, the resulting leading perturbations simplify to
\begin{eqnarray}
 \delta\rho_\pm(\vrr,t)=\frac{\pi}{\nu_c^\pm}\left(a^{(2)}_{{\nu_c^\pm}01}\sin{2\pi x_\parallel}-b^{(2)}_{{\nu_c^\pm}01}\cos{2\pi x_\parallel}\right)\sin{\nu_c^\pm t}+\left(-s^{(2)}_{{\nu_c^\pm}01}\sin{2\pi x_\parallel+t^{(2)}_{{\nu_c^\pm}01}\cos{2\pi x_\parallel}}\right)\cos{\nu_c^\pm t}
\end{eqnarray}
\begin{eqnarray}
 \delta\vj_\pm(\vrr,t)=\frac{1}{2}\left({\bf a}_{{\nu_c^\pm}01}\cos{2\pi x_\parallel}+{\bf b}_{{\nu_c^\pm}01}\sin{2\pi x_\parallel}\right)\cos{\nu_c^\pm t}+\left({\bf s}_{{\nu_c^\pm}01}\cos{2\pi x_\parallel+{\bf t}_{{\nu_c^\pm}01}\sin{2\pi x_\parallel}}\right)\sin{\nu_c^\pm t}
\end{eqnarray}
with ${\bf a}_{{\nu_c^\pm}01}=(a^{(1)}_{{\nu_c^\pm}01},a^{(2)}_{{\nu_c^\pm}02})$, ${\bf b}_{{\nu_c^\pm}01}=(b^{(1)}_{{\nu_c^\pm}01},b^{(2)}_{{\nu_c^\pm}02})$, ${\bf s}_{{\nu_c^\pm}01}=(s^{(1)}_{{\nu_c^\pm}01},s^{(2)}_{{\nu_c^\pm}02})$, ${\bf t}_{{\nu_c^\pm}01}=(t^{(1)}_{{\nu_c^\pm}01},t^{(2)}_{{\nu_c^\pm}02})$ the coefficients of the Fourier series corresponding to that mode. Introducing these perturbations in (\ref{O2flat}) and imposing $O2>0$ \cite{BD2}, we arrive at a relation between the different coefficients, $a_{01}^{(2)}=\pm t_{01}^{(2)}$, $b_{01}^{(2)}=\mp s_{01}^{(2)}$ for $\nu_c^\pm$. As a result, the dominant perturbation of the density profile once the instability is triggered takes the form of a one-dimensional traveling wave
\beq
\delta\rho(x_\parallel,t)=A\sin{\left[2\pi\left(x_\parallel-x_\parallel^0-\frac{q_\parallel\sigma'_0}{\sigma_0} t \right)\right]} \, ,
\label{TWpert}
\eeq
with $A$ and $x_\parallel^0$ two arbitrary constants. 

With this result in mind, we consider now that the relevant density fields well below the instability conserve a traveling-wave structure, i.e. $\rho(\vrr,t)\equiv \omega(\vrr-\vv t)$, with $\vv$ some velocity vector to be determined in the variational problem. Taking now into account the continuity constraint Eq. (\ref{contappA}) we have that $\vnabla_{\vrr'}\cdot \vj(\vrr')=\vv\cdot \vnabla_{\vrr'}\omega(\vrr')$, with the definition $\vrr'=\vrr-\vv t$. Integrating the previous expression leads to
\beq
\vj(\vrr,t)=\vv\omega(\vrr-\vv t)+\bm{\Phi}(\vrr-\vv t) \, ,
\eeq
where $\bm{\Phi}(\vrr-\vv t)$ is an arbitrary divergence-free vector field. To explicitly account for the constraint (\ref{currappA}) on the empirical current, we now split the field $\bm{\Phi}$ into two terms, $\bm{\Phi}(\vrr-\vv t)={\bf k}+\bm{\phi}(\vrr-\vv t)$, where ${\bf k}=\vq-\vv \rho_0$ is a constant vector fixed by constaints (\ref{massappA}) and (\ref{currappA}), and $\bm{\phi}(\vrr-\vv t)$ is now an arbitrary divergence-free field with zero integral, see Eqs. (\ref{TWcond2})-(\ref{TWcond3}) below, defining another degree of freedom (a sort of \emph{gauge field}) to be determined in the variational problem. The resulting traveling-wave form of the current field is
\be
\vj(\vrr,t)=\vq -\vv\left[\rho_0-\omega(\vrr-\vv t)\right]+{\bm{\phi}}(\vrr-\vv t) \, .
\ee
Interestingly, the system uses this kind of \emph{gauge freedom} to optimize a given current fluctuation in the symmetry-broken phase, selecting among all possible \emph{gauges} a particular, non-trivial one which maximizes the probability of this event. This sort of gauge freedom is precisely the key feature responsible of the richness of the fluctuation phase diagram for $d>1$.

In this way, under the above traveling-wave assumptions, the current  LDF of Eq. (\ref{LDFappA}) can now be written, after a change of variables $(\vrr-\vv t) \to \vrr$, as
\be
G(\vq)=-\min_{\omega,{\bm{\phi}},\vv}\int_{\Lambda}d\vrr \, {\cal G}_\vq(\omega,\bm{\phi},\vv) ,
\label{LDF2}
\ee
with the definitions
\ben
{\cal G}_\vq(\omega,\bm{\phi},\vv)&\equiv & \frac{1}{2\sigma(\omega)}\bm{\mathcal{J}}_\vq(\omega,\bm{\phi},\vv) \cdot \mA^{-1}\bm{\mathcal{J}}_\vq(\omega,\bm{\phi},\vv), \\
\bm{\mathcal{J}}_\vq(\omega,\bm{\phi},\vv)&\equiv& \vq -\vv\left[\rho_0-\omega(\vrr)\right]+{\bm{\phi}}(\vrr) +D(\omega)\mA\vnabla\omega-\sigma(\omega)\mA\vE \, ,
\label{currdef}
\een
and with the additional constraints
\begin{eqnarray}
& \label{TWcond1}\displaystyle\rho_0=\int_{\Lambda} \omega(\vrr)\, d\vrr\\
&\label{TWcond2}\displaystyle\int_{\Lambda}{\bm{\phi}}(\vrr)\, d\vrr=0\\
&\label{TWcond3}\displaystyle\vnabla\cdot{\bm{\phi}}(\vrr)=0
\end{eqnarray}
To account for these constraints, we employ the method of Lagrange multipliers. In particular, we write
\be
G(\vq)=-\min_{ \scriptstyle \omega,{\bm{\phi}},\vv \atop \scriptstyle \zeta,{\bm{\kappa}},\Psi}\int_{\Lambda}d\vrr \, \tilde{\cal G}_\vq(\omega,\bm{\phi},\vv,\zeta,{\bm{\kappa}},\Psi) ,
\label{LDF3}
\ee
where the modified functional to minimize is
\beq
\tilde{\cal G}_\vq(\omega,\bm{\phi},\vv,\zeta,{\bm{\kappa}},\Psi) \equiv {\cal G}_\vq(\omega,\bm{\phi},\vv) +\zeta\left[\rho_0- \omega(\vrr)\right]+\bm {\kappa}\cdot {\bm{\phi}}(\vrr)+\Psi(\vrr)\vnabla\cdot{\bm{\phi}}(\vrr) \, ,
\eeq
and $\zeta$, ${\bm{\kappa}}$ and $\Psi(\vrr)$ are the Lagrange multipliers associated to the constraints (\ref{TWcond1}), (\ref{TWcond2}) and (\ref{TWcond3}), respectively. Standard variational calculus shows now that the optimal fields and velocity solution of this complex variational problem, denoted as $\omega_\vq(\vrr)$, $\bm{\phi}_\vq(\vrr)$, and $\vv_\vq$, obey the following system of coupled equations,
\begin{eqnarray}
&\displaystyle\left[\frac{\vv_{\vq}}{\sigma(\omega_{\vq})}-\frac{\sigma'(\omega_{\vq})}{2\sigma(\omega_{\vq})^2}\vj_{\vq}\right]\cdot\mA^{-1}\vj_{\vq}  -\left[\left(\frac{D(\omega_{\vq})^2}{2\sigma(\omega_{\vq})}\right)' \vnabla\omega_{\vq}+\frac{D(\omega_{\vq})^2}{\sigma(\omega_{\vq})}\vnabla\right]\cdot\mA\vnabla\omega_{\vq}+\frac{1}{2}\sigma'(\omega_{\vq})\vE\cdot\mA\vE-\zeta=0
\label{TWPDE1}
\end{eqnarray}
\beq
\label{TWPDE2}D(\omega_{\vq})\vnabla\omega_{\vq}+\mA^{-1}\vj_{\vq}+\sigma(\omega_{\vq})\left[{\bm{\kappa}}-\vnabla\Psi\right]=0 \, ,
\eeq
\beq
\label{TWPDE3}\int_{\Lambda}d\vrr\left(\frac{\omega_{\vq}-\rho_0}{\sigma(\omega_{\vq})}\right) \mA^{-1}\vj_{\vq}=0 \, ,
\eeq
where we have defined $\vj_\vq(\vrr)\equiv \vq -\vv_\vq\left[\rho_0-\omega_\vq(\vrr)\right]+{\bm{\phi}}_\vq(\vrr)$ for simplicity in notation.

As discussed above, our local stability analysis shows that whenever the transition is unleashed, the leading instability is a density wave with  structure in one dimension only, determined either by the minimum-anisotropy direction, see condition (iii) above, or by the orientation of the current vector  for isotropic systems, see (iv). Such a $1d$ traveling wave will dominate the optimal solution of our variational problem \emph{at least} in a finite region below the transition line, so we now assume $1d$ optimal traveling-wave fields of the form $\omega_{\vq}(x_{\parallel})$ and $\bm{\phi}_{\vq}(x_{\parallel})$ (recall that we denote as $x_\parallel$ the direction of structure formation, and $x_\perp$ the orthogonal, structureless direction). Next we decompose the optimal vector field $\bm{\phi}_\vq$ along the $\parallel$- and $\perp$-directions, $\bm{\phi}_\vq(x_{\parallel})=[\phi_\vq^\parallel(x_{\parallel}),\phi_\vq^\perp(x_{\parallel})]$. The divergence-free constraint (\ref{TWcond3}) on $\bm{\phi}_\vq(x_{\parallel})$ immediately implies that $\phi_\vq^\parallel$ is in fact a constant, while the zero-integral constraint (\ref{TWcond2}) sets this constant to zero, resulting in a simplfied form of the vector field $\bm{\phi}_\vq(x_{\parallel})=[0,\phi_\vq^\perp(x_{\parallel})]$. This in turn implies that 
\be
j_\vq^\parallel(x_{\parallel})=q_{\parallel} - v_{\parallel}[\rho_0-\omega_\vq(x_{\parallel})] \, . \nonumber 
\ee
Now, by differentiating the $\perp$-component of Eq. (\ref{TWPDE2}) with respect to $x_{\perp}$, it is straightforward to see that $\partial_{\perp}\Psi$ is a function of $x_{\parallel}$ at most. Moreover, doing the same differentiation on the $\parallel$-component of (\ref{TWPDE2}), we obtain that $\partial_\parallel\partial_{\perp}\Psi = 0$, which together with the previous observation implies that $\partial_{\perp}\Psi$ is indeed a constant. Using this information in the $\perp$-component of Eq. (\ref{TWPDE2}) together with constraint (\ref{currappA}) on the empirical current, we obtain that
\be
j_\vq^\perp(x_{\parallel}) = q_\perp \frac{\sigma[\omega_{\vq}(x_{\parallel})]}{\int_0^1\sigma[\omega_{\vq}(x_{\parallel})] dx_{\parallel}} \, . 
\label{jqperp}
\ee

We next focus on Eq. (\ref{TWPDE1}). Multiplying this equation by $\omega_\vq'(x_\parallel)$, using that $dF[\omega_{\vq}(x_{\parallel})]/dx_{\parallel} = F'(\omega_{\vq})\, \omega_{\vq}'(x_{\parallel})$ for any arbitrary functional $F(\omega_{\vq})$, and the identity 
\be
\frac{d\vj_\vq(x_{\parallel})}{dx_{\parallel}} =\vv_\vq\omega_\vq'(x_{\parallel})+\frac{d{\bm{\phi}}_\vq(x_{\parallel})}{dx_{\parallel}} \, , \nonumber
\ee
Eq. (\ref{TWPDE1}) can be rewritten as
\beq\nonumber
\frac{d}{dx_\parallel}\left[\frac{1}{2\sigma(\omega_\vq)}\vj_\vq\cdot\mA^{-1}\vj_{\vq}-a_{\text{min}}\frac{D(\omega_\vq)^2}{2\sigma(\omega_\vq)}\left(\frac{d\omega_{\vq}}{dx_\parallel}\right)^2+\frac{1}{2}\sigma(\omega_\vq)\vE\cdot\mA\vE\right]-\frac{1}{a_{\text{max}}\sigma(\omega_\vq)}\frac{d\phi_\vq^\perp(x_\parallel)}{dx_\parallel}j_\vq^\perp(x_\parallel)-\zeta\omega_\vq'(x_\parallel)=0. \nonumber
\eeq
Integrating this equation once and taking into account the form of $j_\vq^\perp(x_\parallel)$, see Eq. (\ref{jqperp}), we arrive at a differential equation for the optimal traveling-wave profile
\beq
\label{1dPDE1}X(\omega_{\vq})\left(\frac{d\omega_{\vq}}{dx_\parallel}\right)^2-Y(\omega_{\vq})+\tilde{K}\omega_{\vq}(x_\parallel)-K=0 \, ,
\eeq
with $K$ and $\tilde{K}$ two constants which comprise the Lagrange multiplier $\zeta$, the wave velocity $\vv_\vq$, and information on the boundary conditions, and where we have defined
\beq
X(\omega)\equiv \frac{D(\omega)^2}{2\sigma(\omega)}a_{\text{min}} \, ,
\eeq
\beq
Y(\omega)\equiv \frac{\sigma(\omega)}{2}\left(\vE\cdot\mA\vE+\frac{[q_{\parallel}-v_{\parallel}(\rho_0-\omega)]^2}{a_{\text{min}}\sigma(\omega)^2}-\frac{q_{\perp}^2}{a_{\text{max}}(\int_0^1\sigma(\omega)dx_{\parallel})^2}\right) \, .
\eeq
Finally, two additional equations follow from the $\parallel$-component of Eq. (\ref{TWPDE3}) and constraint (\ref{TWcond1})
\beq
\label{1dPDE2}\int_0^1dx_{\parallel}\frac{[\omega_{\vq}(x_\parallel)-\rho_0]}{a_{\text{min}}\sigma(\omega_{\vq})}\left[q_{\parallel}-v_{\parallel}(\rho_0-\omega_{\vq}(x_\parallel))\right]=0 \, ,
\eeq
\beq
\label{1dPDE3}\rho_0=\int_0^1\omega_{\vq}(x_\parallel) \, dx_{\parallel} \, ,
\eeq
which complete the system of coupled integro-differential equations for the optimal fields.

In order to solve this system, we now introduce a reparametrization which simplifies the numerical evaluation of the optimal $1d$ density wave profile and thus of the current LDF $G(\vq)$. First note that, in our geometry, Eq. (\ref{1dPDE1}) leads to a periodic optimal profile \emph{symmetric} around $x_\parallel=1/2$ (recall that $x_\parallel\in[0,1]$), i.e. with reflection symmetry $x_\parallel\to 1-x_\parallel$. Next we consider the possible maxima and minima of the optimal density wave. For models with a quadratic mobility transport coefficient $\sigma(\omega)$, as the WASEP and KMP models typically studied in literature, the number of possible maxima $\omega_+$ and minima $\omega_-$ of the curve $\omega_\vq(x_\parallel)$ is rather restricted, see Eq. (\ref{1dPDE1}) once particularized for $\omega'_{\vq}(x_\parallel)=0$. In the simplest case \cite{wePRE,weJSP}, a single maximum $\omega_+=\omega_{\vq}(x_{\parallel}^+)$ and minimum $\omega_-=\omega_{\vq}(x_{\parallel}^-)$ will appear, such that the position of two consecutive extrema $x_{\parallel}^+$ and $x_{\parallel}^-$ is such that $|x_{\parallel}^+(k) - x_{\parallel}^-(k)|=1/2n$, with $n$ the number of cycles in the unit interval. One can then study numerically the dependence of the current LDF on the number $n$ of cycles, finding that $n=1$ is the optimal case. We hence restrict hereafter to $1d$ density waves with a single maximum and minimum with $n=1$. As a result, we can express now the constants $\tilde{K}$ and $K$ of Eq. (\ref{1dPDE1}) in terms of these extrema
\be
Y(\omega_\pm)=\tilde{K}\omega_\pm-K \, .
\label{Konstants}
\ee
The values of these extrema $\omega_\pm$ can be obtained from the constraints on the distance between them and the total density of the system. In particular, the first constraint leads to the following equation,
\beq
\label{1dperiod}1=\int_{0}^{1}dx_{\parallel}=2\int_{\omega_-}^{\omega_+}\frac{d\omega_{\vq}}{\omega'_{\vq}}=2\int_{\omega_-}^{\omega_+} {f(\omega_{\vq})}\, d\omega_{\vq}
\eeq
with 
\be
f(\omega_{\vq})\equiv \sqrt{\frac{X(\omega_{\vq})}{Y(\omega_{\vq})-\tilde{K}\omega_{\vq}+K} }
\ee
as derived from Eq. (\ref{1dPDE1}), while the constraint on the total density leads to
\be
\label{1drho0}\rho_0=\int_{0}^{1}\omega_{\vq}(x_{\parallel})\, dx_{\parallel}=2\int_{\omega_-}^{\omega_+}\frac{\omega_{\vq}}{\omega'_{\vq}}d\omega_{\vq}=2\int_{\omega_-}^{\omega_+} {\omega_{\vq} f(\omega_{\vq})}\, d\omega_{\vq} \, .
\ee
Note that the unknown variables $\omega_\pm$ appear as integration limits in Eqs. (\ref{1dperiod}) and (\ref{1drho0}), difficulting the numerical solution of this problem. However, a suitable change of variables in $\omega$-space allows to drop this dependence. In particular, we write now $\omega_{\vq}\equiv\omega_- + \Omega(\omega_+-\omega_-)$, with $\Omega\in[0,1]$, and define $h(\Omega)\equiv (\omega_+-\omega_-) f[\omega_- + \Omega(\omega_+-\omega_-)]$. With this choice, constraints (\ref{1dperiod}) and (\ref{1drho0}), together with Eq. (\ref{1dPDE2}) for the velocity, now read
\beq
\frac{1}{2}=\int_0^1 h(\Omega) \, d\Omega \, ,
\eeq
\beq
\frac{\rho_0}{2}=\int_0^1 \omega_{\vq}(\Omega) h(\Omega) \, d\Omega \, ,
\eeq
\beq
\int_0^1 h(\Omega) \frac{[\omega_{\vq}(\Omega)-\rho_0]}{a_{\text{min}}\sigma[\omega_{\vq}(\Omega)]}\left[q_{\parallel}-v_{\parallel}(\rho_0-\omega_{\vq}(\Omega))\right] \, d\Omega=0 \, .
\eeq
The solution of this three integral equations for a particular model and a given current vector $\vq$ leads to particular values of the parameters $\omega_-$, $\omega_+$ and $v_{\parallel}$, which can be used in turn to obtain the constants $K$ and $\tilde{K}$ from Eq. (\ref{Konstants}) needed to solve numerically the differential equation (\ref{1dPDE1}) for the optimal density wave profile \cite{wePRE,weJSP} and thus obtain the current LDF $G(\vq)$.

A related, interesting function is the dynamical free energy (dFE) $\mu(\vlamb)$ discussed in the main text. This is nothing but the scaled cumulant generating function associated to the current probability distribution $P_{\tau}(\vq)$, defined as $\mu(\vlamb)\equiv \lim_{t\to\infty} t^{-1} \ln \la \text{e}^{t\vlamb\cdot\vq}\ra$ or equivalently as the Legendre transform of the current LDF,
\be
\mu(\vlamb)=\max_{\vq}[G(\vq)+\vlamb\cdot \vq] \, ,
\ee
with $\vlamb$ a vector conjugated to the current. This function can be seen as the conjugate \emph{potential} to $G(\vq)$, a relation equivalent to the free energy being the Legendre transform of the internal energy in thermodynamics. The above MFT analysis of the dynamic phase transition can be developed also in terms of $\mu(\vlamb)$, and this allows a direct comparison with the results of numerical experiments based on the cloning Monte Carlo method, see main text. In particular, defining ${\vz}\equiv \vlamb+\vE$, it can be shown that a line of critical values $\vz_c$ exists at which the instability appears, defined by the equation $\vz_c\cdot\mA\vz_c=\Xi_c$, with $\Xi_c$ the critical threshold defined in Eq. (\ref{xic}) above. This critical line separates a phase of Gaussian current statistics and homogeneous optimal profiles, corresponding to a quadratic dFE $\mu_{\text{G}}(\vz)= \sigma_0(\vz\cdot \mA\vz - \vE\cdot\mA\vE)/2$, see Eq. (\ref{Gqgaus}), and the non-Gaussian, traveling-wave phase. As before, for systems with $\sigma_0''>0$ (as the KMP model) the Gaussian regime dominates for $\vz \cdot\mA\vz\le \Xi_c$ while the traveling-wave region appears for $\vz \cdot\mA\vz > \Xi_c$ and $\forall \vE$. On the other hand, for systems with $\sigma_0''<0$ (as the WASEP studied here) a line of critical values of the external field exist, defined by Eq. (\ref{ec}), beyond which the instability appears, $\vE\cdot\mA\vE \ge |\Sigma_c|$. In this strong field case, Gaussian statistics are expected $\forall \vz$ except for a region defined by $\vz \cdot\mA\vz\le \Xi_c$, where current fluctuations are non-Gaussian. 

\begin{figure}
\vspace{-0.3cm}
\includegraphics[width=15.5cm]{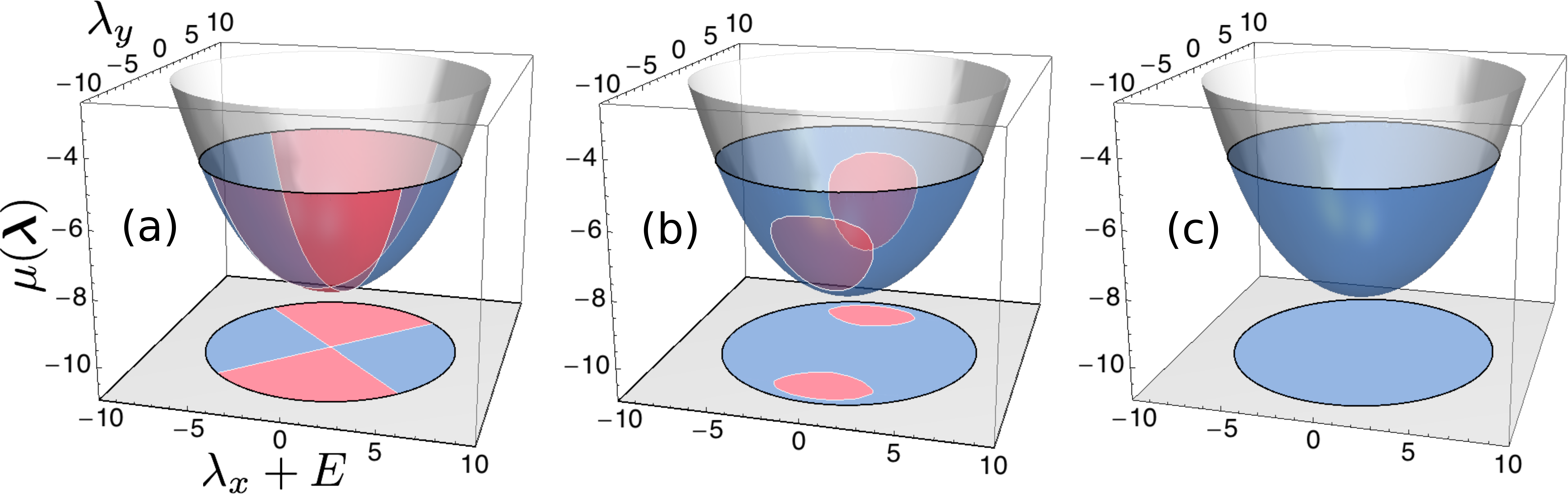}
\vspace{-0.2cm}
\caption{(Color online) Dynamical free energy of the current for the $2d$-WASEP in an external field $\vE=(10,0)$ along the $x$-direction, as derived from MFT in the case of (a) no anisotropy, $\epsilon=0$, (b) mild anisotropy, $0<\epsilon<\epsilon_c$, and (c) strong anisotropy,  $\epsilon>\epsilon_c$. A DPT appears between a Gaussian phase (light gray) with homogeneous optimal pathways, see sketch in Fig. \ref{fig2app}.a representing a typical configuration trajectory in this case, and two different non-Gaussian symmetry-broken phases for low currents characterized by traveling-wave jammed states. The first DPT is $2^{\text{nd}}$-order while the two symmetry-broken phases are separated by lines of $1^{\text{st}}$-order DPTs, see Fig. \ref{fig3app} below.
}
\label{fig1app}
\end{figure}

In this paper we are interested in the current statistics of the $2d$ anisotropic weakly asymmetric simple exclusion process (WASEP), see the main text. At the macroscopic level this model is defined by a diffusivity and mobility matrices $\hD(\rho)= D(\rho)\mA$ and $\hS(\rho)=\sigma(\rho)\mA$, respectively, with $D(\rho)=1/2$ and $\sigma(\rho)=\rho(1-\rho)$ (note that $\sigma''(\rho)<0$). The diagonal anisotropy matrix $\mA$ has components $\mA_{\alpha\beta}=a_\alpha\delta_{\alpha\beta}$, with $\alpha,\beta=x$ or $y$. In particular, we consider systems such that $a_x=1+\epsilon$ and $a_y=1-\epsilon$, with $\epsilon$ an anisotropy parameter. The reason behind this choice is that, for finite lattice systems of moderate size $L$ as the ones we can simulate effectively using the cloning method, a strong external field $\vE$ induces an \emph{effective} anisotropy in the medium, enhancing diffusivity and mobility along the field direction. This effect is modeled in our case, with $\vE$ in the $x$-direction, with a parameter $\epsilon\ge 0$ so that the direction of minimum anisotropy (if any) is $y$. Using these definitions, one can particularize the previous theoretical framework for the $2d$ anisotropic WASEP and proceed to solve numerically the variational problem for the current dFE $\mu(\vlamb)$ and the optimal profiles.

\begin{figure}
\vspace{-0.3cm}
\includegraphics[width=15.5cm]{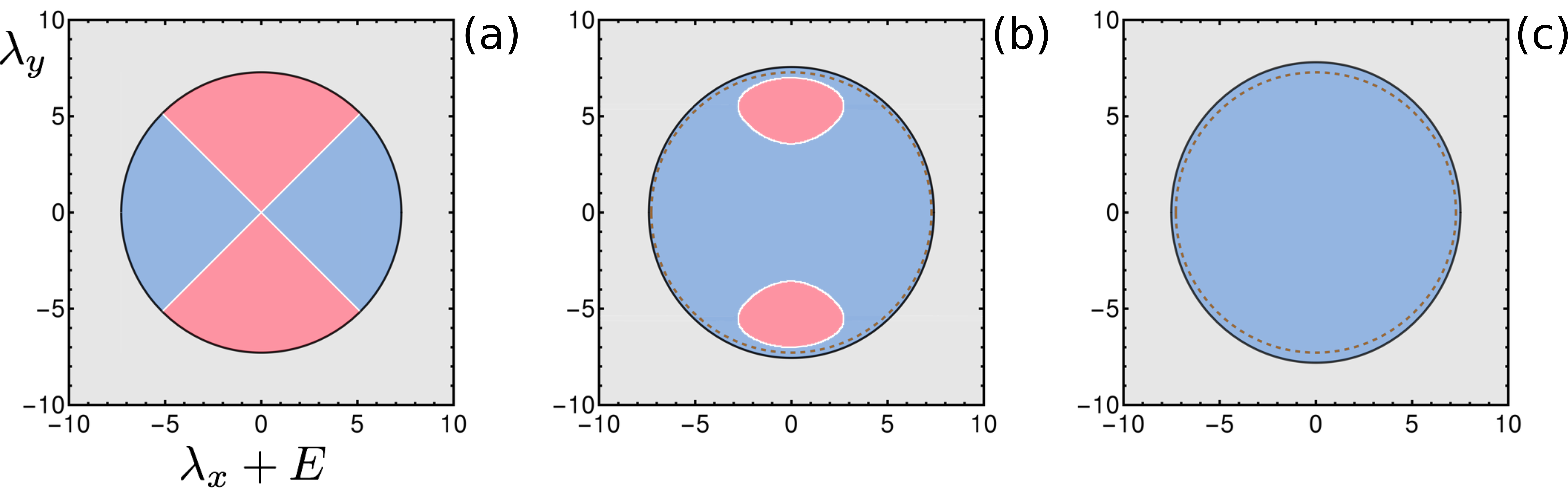}
\vspace{-0.2cm}
\caption{(Color online) A closer look at the phase diagrams for current fluctuations in the case of (a) no anisotropy, $\epsilon=0$, (b) mild anisotropy, $0<\epsilon<\epsilon_c$, and (c) strong anisotropy,  $\epsilon>\epsilon_c$, corresponding to the bottom projections in Fig. \ref{fig1app}. The $2^{\text{nd}}$-order DPT between the Gaussian phase (light gray) and the two different traveling-wave, non-Gaussian phases (dark blue and red) corresponds to the black thick line, while the $1^{\text{st}}$-order DPT separating both symmetry-broken non-Gaussian phases is depicted as a white thin line. Panels (b) and (c) also include a dashed line which corresponds to the $2^{\text{nd}}$-order DPT line for $\epsilon=0$. This shows that the shape of this critical line does change as the anisotropy parameter $\epsilon$ increases.
}
\label{fig1bisapp}
\end{figure}

\begin{figure}
\vspace{-0.3cm}
\includegraphics[width=15.5cm]{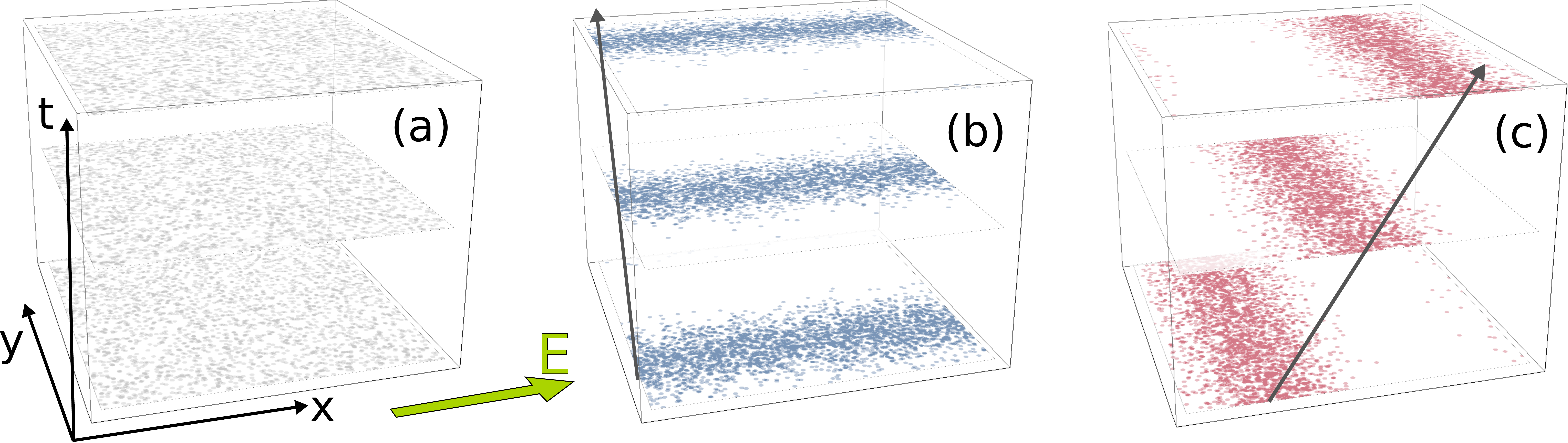}
\vspace{-0.2cm}
\caption{(Color online) Raster plots of typical configuration trajectories for the anisotropic $2d$ WASEP in the Gaussian current fluctuation phase (a), and in the two different non-Gaussian symmetry-broken phases for low currents, (b) and (c), see also Fig. \ref{fig1app}. These two novel phases are characterized by traveling density waves which jam particle flow along the field direction, (b) and blue phase in Fig. \ref{fig1app}, or along the direction orthogonal to $\vE$, (c) and red phase Fig. \ref{fig1app}. 
}
\label{fig2app}
\end{figure}

The solution of this problem shows that the interplay between the external field, the current and the anisotropy leads to a rich phase diagram for current fluctuations. Fig. \ref{fig1app} shows $\mu(\vlamb)$, as derived from our MFT calculations, for three different values of the anisotropy $\epsilon$. In all cases, the dynamic phase transition (DPT) between the Gaussian (light gray) and non-Gaussian (dark colors) phases appears for $\vz_c\cdot\mA\vz_c=\Xi_c$. Fig. \ref{fig1bisapp} shows the phase diagrams for current fluctuations for the different anisotropy parameters (corresponding to the bottom projections of Fig. \ref{fig1app}), and Fig. \ref{fig2app} shows raster plots sketching typical configuration trajectories for WASEP in the Gaussian current fluctuation phase, Fig. \ref{fig2app}.a, and in the two different non-Gaussian symmetry-broken phases which appear for low currents, Figs. \ref{fig2app}.b-c. In general, we find numerically that different traveling wave structures dominate different parts of the symmetry-broken, non-Gaussian phase, see Fig. \ref{fig1app}. For isotropic systems, $\epsilon=0$, the optimal density traveling wave for subcritical vectors $\vz=(z_x,z_y)$ with $|z_x|>|z_y|$ ($|z_x|<|z_y|$) has structure along the $y$-direction ($x$-direction), preserving deep into the non-Gaussian phase the result derived from our local stability analysis right below the transition line, see item (iv) above. On the other hand, for anisotropic systems ($\epsilon>0$) the transition triggers the formation of a density traveling wave with structure only along the minimum anisotropy, $y$-direction, see Figs. \ref{fig1app}.b-c, \ref{fig1bisapp}.b-c and \ref{fig2app}.b, in agreement with item (iii) above. However, for mild anisotropy we find deep into the non-Gaussian regime two pockets of the second symmetry-broken phase, i.e. the one with structure along the maximum anisotropy axis, see Figs. \ref{fig1app}.b, \ref{fig1bisapp}.b and \ref{fig2app}.c. These two patches decrease with increasing $\epsilon$, up to a critical anisotropy $\epsilon_c\approx 0.035$ beyond which only the minimum-anisotropy density wave appears in the non-Gaussian regime, see Figs. \ref{fig1app}.c and \ref{fig1bisapp}.c. 

\begin{figure}
\vspace{-0.3cm}
\includegraphics[width=14.5cm]{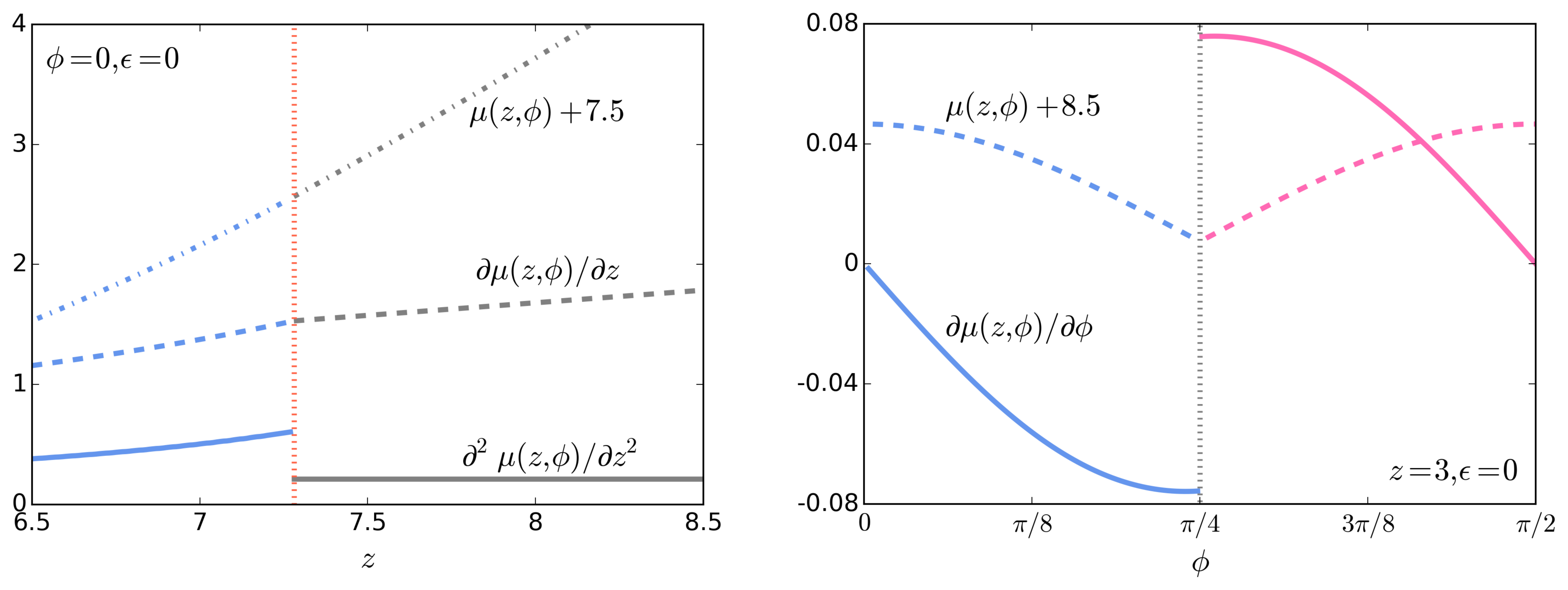}
\vspace{-0.2cm}
\caption{(Color online) Left: Dynamical free energy for the current $\mu(z,\phi)$, with $\vz=\vlamb+\vE$, as a function of $z=|\vz|$ for $\phi=0$ in the isotropic case ($\epsilon=0$), see Fig. \ref{fig1app}.a, as well as its first and second partial derivative with respect to $z$. Note that $\mu(z,\phi)$ has been shifted vertically for the sake of clarity. The vertical dotted line signals the DPT between the Gaussian, homogeneous current fluctuation phase ($z>z_c(\phi)$) and the non-Gaussian, symmetry broken phase ($z<z_c(\phi)$) with jammed density waves along the field direction, see Fig. \ref{fig2app}.b. The dynamical free energy exhibits a kink in its first derivative and an associated discontinuity in the second derivative, a hallmark of a second-order phase transition. Similar discontinuities in $\partial^2_z\mu(z,\phi)$ appear at $z_c(\phi)$ $\forall \phi\in[0,2\pi]$. Right: $\mu(z,\phi)$ vs $\phi$ for $\phi\in [0,\pi/2]$ and $z=3$ in the isotropic case ($\epsilon=0$), see Fig. \ref{fig1app}.a, as well as its first derivative with respect to $\phi$. As before, $\mu(z,\phi)$ has been shifted vertically for clarity. The vertical dotted line signals the DPT separating the two distinct non-Gaussian symmetry-broken phases with jammed states along the field direction ($\phi<\pi/4$) or orthogonal to it ($\phi>\pi/4$). While $\mu(z=3,\phi)$ is continuous across the transition, it exhibits a kink at $\phi_c=\pi/4$ and an associated discontinuity in $\partial_\phi\mu(z=3,\phi)$, signaling the first-order character of this DPT between the two symmetry-broken non-Gaussian phases.
}
\label{fig3app}
\end{figure}

Next, we investigate the order of the different DPT's showing up in the current statistics of this model. We first focus on the DPT from the Gaussian to the non-Gaussian phase at $\vz_c\cdot\mA\vz_c=\Xi_c$. Left panel in Fig. \ref{fig3app} shows $\mu(\vz)$ as a function of $z=|\vz|$ for a current angle $\phi=0$ in the isotropic case ($\epsilon=0$), as well as its first and second partial derivatives with respect to $z$ at constant $\phi$. Clearly, the dynamical free energy exhibits a kink in its first derivative and a related discontinuity in the second derivative, a hallmark of a second-order phase transition. Similar discontinuities in $\partial^2_z\mu(z,\phi)$ appear at $z_c(\phi)$ $\forall \phi\in[0,2\pi]$. Therefore, as happens also in the simpler DPT's already described and observed in $1d$ oversimplified transport models \cite{weSSB,wePRE}, the DPT from the Gaussian, homogeneous phase and the non-Gaussian, traveling-wave phases is of second order type. 

On the other hand, the DPT between different symmetry-broken phases for $\vz_c\cdot\mA\vz_c<\Xi_c$ and mild or no anisotropy, see Fig. \ref{fig1app}.a-b, is clearly discontinuous. Indeed, right panel in Fig. \ref{fig3app} shows $\mu(\vz)$ as a function of the angle $\phi\in [0,\pi/2]$ for $z=3$ (deep into the symmetry-broken phase) in the isotropic case $\epsilon=0$, see Fig. \ref{fig1app}.a, as well as its first derivative with respect to $\phi$ at constant $z$. The vertical dotted line in this plot signals the DPT separating the two distinct non-Gaussian symmetry-broken phases with traveling jammed states along the field direction ($\phi<\pi/4$) or orthogonal to it ($\phi>\pi/4$). While $\mu(z=3,\phi)$ is continuous across the transition, it exhibits a kink at $\phi_c=\pi/4$ and an associated discontinuity in $\partial_\phi\mu(z=3,\phi)$, signaling the first-order character of this DPT between the two symmetry-broken non-Gaussian phases. Something similar happens for all other subcritical $z$ and $\epsilon<\epsilon_c$. Interestingly, along these $1^{\text{st}}$-order DPT lines, both traveling wave solutions are equally probable, giving rise to a \emph{coexistence} of two different dynamic fluctuating phases very much reminiscent of standard first-order critical phenomena. 

To end this section we note that, even though our local stability analysis shows that the dominant perturbations immediately beyond the instability line are one-dimensional traveling waves, in principle one could expect more complex two-dimensional (traveling-wave) patterns to emerge deeper into the symmetry-broken phase. In this case, the equations defining the form of the optimal profiles are partial differential equations, see e.g. Eq. (\ref{TWPDE1}), and the uniqueness of their solution is in general unknown. However, one can find some particular solutions which are \emph{local} maximizers of the MFT action for currents. The particular $2d$ solutions we have explored numerically do not improve the current LDF when compared to their $1d$ counterparts described above. In any case, we cannot discard exotic $2d$ solutions not yet explored, though our simulation results in the main text strongly support that $1d$ traveling waves are the global optimal solutions in all cases.

\begin{figure}
\vspace{-0.3cm}
\includegraphics[width=11.5cm]{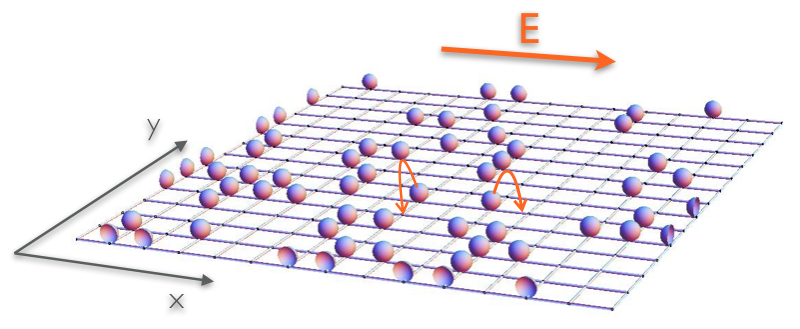}
\vspace{-0.2cm}
\caption{(Color online) Sketch of the $2d$ weakly asymmetric simple exclusion process (WASEP). $M$ particles evolve in a $2d$ square lattice of size $N=L\times L$ with periodic boundary conditions, such that $M\le N$ and the global density is $\rho_0=M/N$. Each site might be occupied by one particle at most, which jumps stochastically to neighboring empty sites at a rate $r^\alpha_{\pm}\equiv\text{exp}[\pm E_\alpha/L]/2$ for moves along the $\pm{\alpha}$-direction, $\alpha=x,y$, with $\vE=(E_x,E_y)$ the external field.
}
\label{fig4app}
\end{figure}

\section{An order parameter for the dynamic phase transition}

In this section we describe in more detail the novel order parameter introduced in the main text to detect and characterize the onset of the $2^{\text{nd}}$-order DPT predicted by MFT. Let us first fix some notation. The $2d$-WASEP is defined at the microscopic level on a $2d$ square lattice of size $N=L\times L$ with periodic boundaries where $M\leq N$ particles evolve, so the global density is $\rho_0 = M/N$, see sketch in Fig. \ref{fig4app}. Each lattice site may contain at most one particle, so the state of the system is defined by an occupation vector ${\bf n}\equiv\{n_{ij}=0,1;i,j\in [1,L] \}$, with $M=\sum_{i,j=1}^{L}n_{ij}$. Particles perform stochastic jumps to neighboring empty sites at a rate $r^\alpha_{\pm}\equiv\text{exp}[\pm E_\alpha/L]/2$ for jumps along the $\pm{\alpha}$-direction, $\alpha=x,y$, with $\vE=(E_x,E_y)$ the external field.

\begin{figure}
\vspace{-0.3cm}
\includegraphics[width=14.5cm]{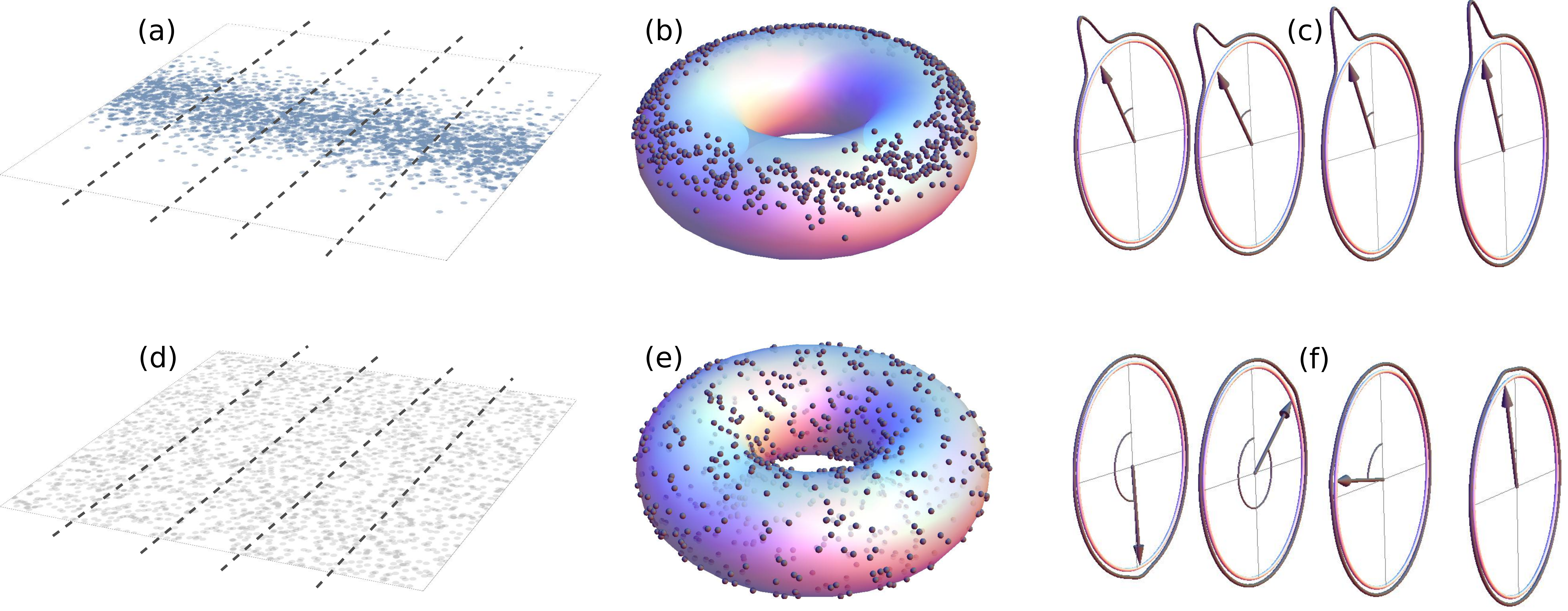}
\vspace{-0.2cm}
\caption{(Color online) Tomographic analysis to define an order parameter for the DPT. Order is expected to emerge across the DPT in the form of $1d$ coherent traveling waves (a) which jam particle flow along one direction. To detect these jams, we slice microscopic configurations along principal axes (see dashed lines in (a)). Due to the periodic boundaries, the systemÕs topology is in fact that of a torus, as in (b), so each slice can be considered as a 1d ring of fixed radius embedded in 2d, with a given angular mass distribution (c) depending on the positions of the particles in the slice. A small dispersion  $\sigma_x^2$ of the angular centers of mass across the different slices, (c), will signal the formation of a coherent jam along the $x$-direction and the associated density wave in the orthogonal direction, see (a). A similar analysis in the homogeneous, Gaussian phase leads to a typically large dispersion $\sigma_x^2$, see (d)-(f).
}
\label{fig5app}
\end{figure}

As described in the previous section, macroscopic fluctuation theory predicts a dynamic phase transition in the current statistics of this model, for currents well below the average. In particular, we expect order to emerge across the DPT in the form of $1d$ coherent traveling waves which jam particle flow along one direction, thus facilitating low-current deviations. The interplay described above among the external field, anisotropy and currents opens the door to different, competing symmetry-broken phases, see Figs. \ref{fig1app}, \ref{fig1bisapp} and \ref{fig2app}, and our aim here is to determine which ones do emerge in our simulations. To define an appropriate order parameter we perform now a \emph{tomographic analysis} by taking $1d$ sections of our $2d$ system. In particular we consider a microscopic particle configuration ${\bf n}$ and slice it along one of the principal axes, say $x$, defining the $j$-slice configuration ${\bf n}_j\equiv\{n_{ij};i\in [1,L] \}$, with $M_j=\sum_{i=1}^{L}n_{ij}$ the total number of particles in this slice and $M=\sum_{j=1}^{L}M_j$, see e.g Figs. \ref{fig5app}.a,d. To properly take into account the periodic boundaries (i.e. the system torus topology, see Figs. \ref{fig5app}.b,e), we consider each $j$-slice as a $1d$ ring of fixed radius embedded in $2d$ where each site $i\in[1,L]$ is assigned an angle $\theta_i=2\pi i/L$, and compute the angular position of the center of mass for the $j$-slice, $\theta_{\text{cm}}^{(j)}$. This is defined as
\be
\theta_{\text{cm}}^{(j)}\equiv \tan^{-1}(\frac{S_j}{C_j})
\ee
with the additional definitions
\ben
S_j &\equiv& \frac{1}{M_j}\sum_{i=1}^Ln_{ij}\sin\theta_i \, , \\
C_j &\equiv& \frac{1}{M_j}\sum_{i=1}^Ln_{ij}\cos\theta_i \, .
\een
Clearly, a small dispersion of the angular centers of mass across the different slices will signal the formation of a coherent jam along the $x$-direction and the associated density wave in the orthogonal direction, see Fig. \ref{fig5app}.c. On the other hand, a large dispersion of $\theta_{\text{cm}}^{(j)}$ across the different $j\in[1,L]$ is the typical signature of a structureless, homogeneous random configuration, see Figs. \ref{fig5app}.d,f. In this way, we write 
\be
\sigma_x^2\equiv \la(\theta_{\text{cm}}^{(j)})^2\ra_x - \la{\theta_{\text{cm}}^{(j)}}\ra_x^2 \, ,
\ee
where we have defined
\be
\la f_j\ra_x \equiv \frac{1}{L}\sum_{j=1}^L f_j \, ,
\ee
for any arbitrary local observable $f_j$, and define the \emph{tomographic $x$-coherence} as 
\be
\Delta_x(\vlamb)\equiv 1-\la\sigma_x^2\ra_\vlamb \, ,
\ee
where the average $\la\cdot\ra_\vlamb$ is taken over the biased $\vlamb$-ensemble, i.e. over all trajectories statistically relevant for a rare event of fixed $\vlamb$ \cite{Derrida1,weJSP,BD2}. We can define in an equivalent way the tomographic $y$-coherence $\Delta_y(\vlamb)$ to detect particle jams along the $y$-direction, and Fig. 3.d in the main text shows these two order parameters measured across the DPT as a function of $z= |\vz|$, with $\vz\equiv\vlamb + \vE$. 

Remarkably, $\Delta_x(z)$ increases steeply for $\vz\cdot\mA\vz\le \Xi_c$ and \emph{all angles} $\phi$ of the current vector, while $\Delta_y(z)$ remains small and does not change appreciably across the DPT, clearly indicating that a coherent particle jam emerges along the $x$-direction in all cases, as in the sketch of Fig. \ref{fig5app}.a above. This means that only one of the two possible symmetry-broken phases appear in our simulations (regardless of the current vector orientation), as expected from MFT in the supercritical anisotropy regime $\epsilon>\epsilon_c$, see Fig. \ref{fig1app}.c, and consistent with the measured effective anisotropy $\epsilon\approx0.038>\epsilon_c$, see inset in Fig. 3.d of the main text. Note also that the behavior of both $\Delta_\alpha$ ($\alpha=x,y$) across the DPT is consistent with the emergence of a traveling wave with structure in $1d$ and not in $2d$, as in the latter case both $\Delta_\alpha$ should increase upon crossing $z_c(\phi)$. Moreover, the acute but continuous change of $\Delta_x(\vz)$ across the DPT is consistent with a second-order transition, in agreement with the MFT prediction.


\begin{thebibliography}{62}%
\makeatletter
\providecommand \@ifxundefined [1]{%
 \@ifx{#1\undefined}
}%
\providecommand \@ifnum [1]{%
 \ifnum #1\expandafter \@firstoftwo
 \else \expandafter \@secondoftwo
 \fi
}%
\providecommand \@ifx [1]{%
 \ifx #1\expandafter \@firstoftwo
 \else \expandafter \@secondoftwo
 \fi
}%
\providecommand \natexlab [1]{#1}%
\providecommand \enquote  [1]{``#1''}%
\providecommand \bibnamefont  [1]{#1}%
\providecommand \bibfnamefont [1]{#1}%
\providecommand \citenamefont [1]{#1}%
\providecommand \href@noop [0]{\@secondoftwo}%
\providecommand \href [0]{\begingroup \@sanitize@url \@href}%
\providecommand \@href[1]{\@@startlink{#1}\@@href}%
\providecommand \@@href[1]{\endgroup#1\@@endlink}%
\providecommand \@sanitize@url [0]{\catcode `\\12\catcode `\$12\catcode
  `\&12\catcode `\#12\catcode `\^12\catcode `\_12\catcode `\%12\relax}%
\providecommand \@@startlink[1]{}%
\providecommand \@@endlink[0]{}%
\providecommand \url  [0]{\begingroup\@sanitize@url \@url }%
\providecommand \@url [1]{\endgroup\@href {#1}{\urlprefix }}%
\providecommand \urlprefix  [0]{URL }%
\providecommand \Eprint [0]{\href }%
\providecommand \doibase [0]{http://dx.doi.org/}%
\providecommand \selectlanguage [0]{\@gobble}%
\providecommand \bibinfo  [0]{\@secondoftwo}%
\providecommand \bibfield  [0]{\@secondoftwo}%
\providecommand \translation [1]{[#1]}%
\providecommand \BibitemOpen [0]{}%
\providecommand \bibitemStop [0]{}%
\providecommand \bibitemNoStop [0]{.\EOS\space}%
\providecommand \EOS [0]{\spacefactor3000\relax}%
\providecommand \BibitemShut  [1]{\csname bibitem#1\endcsname}%
\let\auto@bib@innerbib\@empty
\bibitem [{\citenamefont {Binney}\ \emph {et~al.}(1992)\citenamefont {Binney},
  \citenamefont {Dowrick}, \citenamefont {Fisher},\ and\ \citenamefont
  {Newman}}]{binney92a}%
  \BibitemOpen
  \bibfield  {author} {\bibinfo {author} {\bibfnamefont {J.~J.}\ \bibnamefont
  {Binney}}, \bibinfo {author} {\bibfnamefont {N.~J.}\ \bibnamefont {Dowrick}},
  \bibinfo {author} {\bibfnamefont {A.~J.}\ \bibnamefont {Fisher}}, \ and\
  \bibinfo {author} {\bibfnamefont {M.}~\bibnamefont {Newman}},\ }\href@noop {}
  {\emph {\bibinfo {title} {The Theory of Critical Phenomena: An Introduction
  to the Renormalization Group}}}\ (\bibinfo  {publisher} {Oxford University
  Press, Inc.},\ \bibinfo {address} {New York, NY, USA},\ \bibinfo {year}
  {1992})\BibitemShut {NoStop}%
\bibitem [{\citenamefont {Zinn-Justin}(2002)}]{zinn-justin02a}%
  \BibitemOpen
  \bibfield  {author} {\bibinfo {author} {\bibfnamefont {J.}~\bibnamefont
  {Zinn-Justin}},\ }\href@noop {} {\emph {\bibinfo {title} {Quantum Field
  Theory and Critical Phenomena; 4th ed.}}},\ Internat. Ser. Mono. Phys.\
  (\bibinfo  {publisher} {Clarendon Press},\ \bibinfo {address} {Oxford},\
  \bibinfo {year} {2002})\BibitemShut {NoStop}%
\bibitem [{\citenamefont {Bertini}\ \emph {et~al.}(2015)\citenamefont
  {Bertini}, \citenamefont {Sole}, \citenamefont {Gabrielli}, \citenamefont
  {Jona-Lasinio},\ and\ \citenamefont {Landim}}]{bertini15a}%
  \BibitemOpen
  \bibfield  {author} {\bibinfo {author} {\bibfnamefont {L.}~\bibnamefont
  {Bertini}}, \bibinfo {author} {\bibfnamefont {A.~De}\ \bibnamefont {Sole}},
  \bibinfo {author} {\bibfnamefont {D.}~\bibnamefont {Gabrielli}}, \bibinfo
  {author} {\bibfnamefont {G.}~\bibnamefont {Jona-Lasinio}}, \ and\ \bibinfo
  {author} {\bibfnamefont {C.}~\bibnamefont {Landim}},\ }\bibfield  {title}
  {\enquote {\bibinfo {title} {Macroscopic fluctuation theory},}\ }\href
  {http://journals.aps.org/rmp/abstract/10.1103/RevModPhys.87.593} {\bibfield
  {journal} {\bibinfo  {journal} {Rev. Mod. Phys.}\ }\textbf {\bibinfo {volume}
  {87}},\ \bibinfo {pages} {593--636} (\bibinfo {year} {2015})}\BibitemShut
  {NoStop}%
\bibitem [{\citenamefont {Derrida}()}]{derrida07a}%
  \BibitemOpen
  \bibfield  {author} {\bibinfo {author} {\bibfnamefont {B.}~\bibnamefont
  {Derrida}},\ }\bibfield  {title} {\enquote {\bibinfo {title} {Non-equilibrium
  steady states: fluctuations and large deviations of the density and of the
  current},}\ }\href
  {http://iopscience.iop.org/article/10.1088/1742-5468/2007/07/P07023}
  {\bibinfo  {journal} {J. Stat. Mech. P07023 (2007)}\ }\BibitemShut {NoStop}%
\bibitem [{\citenamefont {Bodineau}\ and\ \citenamefont
  {Derrida}(2005)}]{bodineau05a}%
  \BibitemOpen
\bibfield  {journal} {  }\bibfield  {author} {\bibinfo {author} {\bibfnamefont
  {T.}~\bibnamefont {Bodineau}}\ and\ \bibinfo {author} {\bibfnamefont
  {B.}~\bibnamefont {Derrida}},\ }\bibfield  {title} {\enquote {\bibinfo
  {title} {Distribution of current in nonequilibrium diffusive systems and
  phase transitions},}\ }\href
  {http://journals.aps.org/pre/abstract/10.1103/PhysRevE.72.066110} {\bibfield
  {journal} {\bibinfo  {journal} {Phys. Rev. E}\ }\textbf {\bibinfo {volume}
  {72}},\ \bibinfo {pages} {066110} (\bibinfo {year} {2005})}\BibitemShut
  {NoStop}%
\bibitem [{\citenamefont {Bertini}\ \emph {et~al.}(2006)\citenamefont
  {Bertini}, \citenamefont {Sole}, \citenamefont {Gabrielli}, \citenamefont
  {Jona-Lasinio},\ and\ \citenamefont {Landim}}]{bertini06a}%
  \BibitemOpen
  \bibfield  {author} {\bibinfo {author} {\bibfnamefont {L.}~\bibnamefont
  {Bertini}}, \bibinfo {author} {\bibfnamefont {A.~De}\ \bibnamefont {Sole}},
  \bibinfo {author} {\bibfnamefont {D.}~\bibnamefont {Gabrielli}}, \bibinfo
  {author} {\bibfnamefont {G.}~\bibnamefont {Jona-Lasinio}}, \ and\ \bibinfo
  {author} {\bibfnamefont {C.}~\bibnamefont {Landim}},\ }\bibfield  {title}
  {\enquote {\bibinfo {title} {Nonequilibrium current fluctuations in
  stochastic lattice gases},}\ }\href
  {http://link.springer.com/article/10.1007/s10955-006-9056-4} {\bibfield
  {journal} {\bibinfo  {journal} {J. Stat. Phys.}\ }\textbf {\bibinfo {volume}
  {123}},\ \bibinfo {pages} {237--276} (\bibinfo {year} {2006})}\BibitemShut
  {NoStop}%
\bibitem [{\citenamefont {Lecomte}\ \emph {et~al.}(2007)\citenamefont
  {Lecomte}, \citenamefont {T{\"a}uber},\ and\ \citenamefont {van
  Wijland}}]{lecomte07b}%
  \BibitemOpen
  \bibfield  {author} {\bibinfo {author} {\bibfnamefont {V.}~\bibnamefont
  {Lecomte}}, \bibinfo {author} {\bibfnamefont {U.C.}\ \bibnamefont
  {T{\"a}uber}}, \ and\ \bibinfo {author} {\bibfnamefont {F.}~\bibnamefont {van
  Wijland}},\ }\bibfield  {title} {\enquote {\bibinfo {title} {Current
  distribution in systems with anomalous diffusion: renormalization group
  approach},}\ }\href
  {http://iopscience.iop.org/article/10.1088/1751-8113/40/7/003/meta}
  {\bibfield  {journal} {\bibinfo  {journal} {J. Phys. A}\ }\textbf {\bibinfo
  {volume} {40}},\ \bibinfo {pages} {1447} (\bibinfo {year}
  {2007})}\BibitemShut {NoStop}%
\bibitem [{\citenamefont {Bodineau}\ \emph {et~al.}(2008)\citenamefont
  {Bodineau}, \citenamefont {Derrida},\ and\ \citenamefont
  {Lebowitz}}]{bodineau08a}%
  \BibitemOpen
  \bibfield  {author} {\bibinfo {author} {\bibfnamefont {T.}~\bibnamefont
  {Bodineau}}, \bibinfo {author} {\bibfnamefont {B.}~\bibnamefont {Derrida}}, \
  and\ \bibinfo {author} {\bibfnamefont {J.L.}\ \bibnamefont {Lebowitz}},\
  }\bibfield  {title} {\enquote {\bibinfo {title} {Vortices in the
  two-dimensional simple exclusion process},}\ }\href
  {http://link.springer.com/article/10.1007/s10955-008-9518-y} {\bibfield
  {journal} {\bibinfo  {journal} {J. Stat. Phys.}\ }\textbf {\bibinfo {volume}
  {131}},\ \bibinfo {pages} {821} (\bibinfo {year} {2008})}\BibitemShut
  {NoStop}%
\bibitem [{\citenamefont {Hurtado}\ and\ \citenamefont
  {Garrido}(2011)}]{hurtado11a}%
  \BibitemOpen
  \bibfield  {author} {\bibinfo {author} {\bibfnamefont {P.~I.}\ \bibnamefont
  {Hurtado}}\ and\ \bibinfo {author} {\bibfnamefont {P.~L.}\ \bibnamefont
  {Garrido}},\ }\bibfield  {title} {\enquote {\bibinfo {title} {Spontaneous
  symmetry breaking at the fluctuating level},}\ }\href
  {http://journals.aps.org/prl/abstract/10.1103/PhysRevLett.107.180601}
  {\bibfield  {journal} {\bibinfo  {journal} {Phys. Rev. Lett.}\ }\textbf
  {\bibinfo {volume} {107}},\ \bibinfo {pages} {180601} (\bibinfo {year}
  {2011})}\BibitemShut {NoStop}%
\bibitem [{\citenamefont {P\'erez-Espigares}\ \emph {et~al.}(2013)\citenamefont
  {P\'erez-Espigares}, \citenamefont {Garrido},\ and\ \citenamefont
  {Hurtado}}]{perez-espigares13a}%
  \BibitemOpen
  \bibfield  {author} {\bibinfo {author} {\bibfnamefont {C.}~\bibnamefont
  {P\'erez-Espigares}}, \bibinfo {author} {\bibfnamefont {P.~L.}\ \bibnamefont
  {Garrido}}, \ and\ \bibinfo {author} {\bibfnamefont {P.~I.}\ \bibnamefont
  {Hurtado}},\ }\bibfield  {title} {\enquote {\bibinfo {title} {Dynamical phase
  transition for current statistics in a simple driven diffusive system},}\
  }\href {http://journals.aps.org/pre/abstract/10.1103/PhysRevE.87.032115}
  {\bibfield  {journal} {\bibinfo  {journal} {Phys. Rev. E}\ }\textbf {\bibinfo
  {volume} {87}},\ \bibinfo {pages} {032115} (\bibinfo {year}
  {2013})}\BibitemShut {NoStop}%
\bibitem [{\citenamefont {Hurtado}\ \emph {et~al.}(2014)\citenamefont
  {Hurtado}, \citenamefont {Espigares}, \citenamefont {del Pozo},\ and\
  \citenamefont {Garrido}}]{hurtado14a}%
  \BibitemOpen
  \bibfield  {author} {\bibinfo {author} {\bibfnamefont {P.~I.}\ \bibnamefont
  {Hurtado}}, \bibinfo {author} {\bibfnamefont {C.~P.}\ \bibnamefont
  {Espigares}}, \bibinfo {author} {\bibfnamefont {J.~J.}\ \bibnamefont {del
  Pozo}}, \ and\ \bibinfo {author} {\bibfnamefont {P.~L.}\ \bibnamefont
  {Garrido}},\ }\bibfield  {title} {\enquote {\bibinfo {title} {Thermodynamics
  of currents in nonequilibrium diffusive systems: theory and simulation},}\
  }\href {http://link.springer.com/article/10.1007/s10955-013-0894-6}
  {\bibfield  {journal} {\bibinfo  {journal} {J. Stat. Phys.}\ }\textbf
  {\bibinfo {volume} {154}},\ \bibinfo {pages} {214--264} (\bibinfo {year}
  {2014})}\BibitemShut {NoStop}%
\bibitem [{\citenamefont {Vaikuntanathan}\ \emph {et~al.}(2014)\citenamefont
  {Vaikuntanathan}, \citenamefont {Gingrich},\ and\ \citenamefont
  {Geissler}}]{vaikuntanathan14a}%
  \BibitemOpen
  \bibfield  {author} {\bibinfo {author} {\bibfnamefont {S.}~\bibnamefont
  {Vaikuntanathan}}, \bibinfo {author} {\bibfnamefont {T.~R.}\ \bibnamefont
  {Gingrich}}, \ and\ \bibinfo {author} {\bibfnamefont {P.~L.}\ \bibnamefont
  {Geissler}},\ }\bibfield  {title} {\enquote {\bibinfo {title} {Dynamic phase
  transitions in simple driven kinetic networks},}\ }\href
  {http://journals.aps.org/pre/abstract/10.1103/PhysRevE.89.062108} {\bibfield
  {journal} {\bibinfo  {journal} {Phys. Rev. E}\ }\textbf {\bibinfo {volume}
  {89}},\ \bibinfo {pages} {062108} (\bibinfo {year} {2014})}\BibitemShut
  {NoStop}%
\bibitem [{\citenamefont {Jack}\ \emph {et~al.}(2015)\citenamefont {Jack},
  \citenamefont {Thompson},\ and\ \citenamefont {Sollich}}]{jack15a}%
  \BibitemOpen
  \bibfield  {author} {\bibinfo {author} {\bibfnamefont {R.~L.}\ \bibnamefont
  {Jack}}, \bibinfo {author} {\bibfnamefont {I.~R.}\ \bibnamefont {Thompson}},
  \ and\ \bibinfo {author} {\bibfnamefont {P.}~\bibnamefont {Sollich}},\
  }\bibfield  {title} {\enquote {\bibinfo {title} {Hyperuniformity and phase
  separation in biased ensembles of trajectories for diffusive systems},}\
  }\href {http://journals.aps.org/prl/abstract/10.1103/PhysRevLett.114.060601}
  {\bibfield  {journal} {\bibinfo  {journal} {Phys. Rev. Lett.}\ }\textbf
  {\bibinfo {volume} {114}},\ \bibinfo {pages} {060601} (\bibinfo {year}
  {2015})}\BibitemShut {NoStop}%
\bibitem [{\citenamefont {Shpielberg}\ and\ \citenamefont
  {Akkermans}(2016)}]{shpielberg16a}%
  \BibitemOpen
  \bibfield  {author} {\bibinfo {author} {\bibfnamefont {O.}~\bibnamefont
  {Shpielberg}}\ and\ \bibinfo {author} {\bibfnamefont {E.}~\bibnamefont
  {Akkermans}},\ }\bibfield  {title} {\enquote {\bibinfo {title} {Le
  {C}hatelier principle for out-of-equilibrium and boundary-driven systems:
  Application to dynamical phase transitions},}\ }\href
  {http://dx.doi.org/10.1103/PhysRevLett.116.240603} {\bibfield  {journal}
  {\bibinfo  {journal} {Phys. Rev. Lett.}\ }\textbf {\bibinfo {volume} {116}}
  (\bibinfo {year} {2016})}\BibitemShut {NoStop}%
\bibitem [{\citenamefont {Zarfaty}\ and\ \citenamefont
  {Meerson}()}]{zarfaty16a}%
  \BibitemOpen
  \bibfield  {author} {\bibinfo {author} {\bibfnamefont {L.}~\bibnamefont
  {Zarfaty}}\ and\ \bibinfo {author} {\bibfnamefont {B.}~\bibnamefont
  {Meerson}},\ }\bibfield  {title} {\enquote {\bibinfo {title} {Statistics of
  large currents in the {Kipnis-Marchioro-Presutti} model in a ring
  geometry},}\ }\href
  {http://iopscience.iop.org/article/10.1088/1742-5468/2016/03/033304/meta}
  {\bibinfo  {journal} {J. Stat. Mech. P033304 (2016)}\ }\BibitemShut {NoStop}%
\bibitem [{\citenamefont {Baek}\ \emph {et~al.}(2017)\citenamefont {Baek},
  \citenamefont {Kafri},\ and\ \citenamefont {Lecomte}}]{baek17a}%
  \BibitemOpen
\bibfield  {journal} {  }\bibfield  {author} {\bibinfo {author} {\bibfnamefont
  {Y.}~\bibnamefont {Baek}}, \bibinfo {author} {\bibfnamefont {Y.}~\bibnamefont
  {Kafri}}, \ and\ \bibinfo {author} {\bibfnamefont {V.}~\bibnamefont
  {Lecomte}},\ }\bibfield  {title} {\enquote {\bibinfo {title} {Dynamical
  symmetry breaking and phase transitions in driven diffusive systems},}\
  }\href {http://journals.aps.org/prl/abstract/10.1103/PhysRevLett.118.030604}
  {\bibfield  {journal} {\bibinfo  {journal} {Phys. Rev. Lett.}\ }\textbf
  {\bibinfo {volume} {118}},\ \bibinfo {pages} {030604} (\bibinfo {year}
  {2017})}\BibitemShut {NoStop}%
\bibitem [{\citenamefont {Karevski}\ and\ \citenamefont
  {Sch\"utz}(2017)}]{karevski17a}%
  \BibitemOpen
  \bibfield  {author} {\bibinfo {author} {\bibfnamefont {D.}~\bibnamefont
  {Karevski}}\ and\ \bibinfo {author} {\bibfnamefont {G.M.}\ \bibnamefont
  {Sch\"utz}},\ }\bibfield  {title} {\enquote {\bibinfo {title} {Conformal
  invariance in driven diffusive systems at high currents},}\ }\href
  {http://journals.aps.org/prl/abstract/10.1103/PhysRevLett.118.030601}
  {\bibfield  {journal} {\bibinfo  {journal} {Phys. Rev. Lett.}\ }\textbf
  {\bibinfo {volume} {118}} (\bibinfo {year} {2017})}\BibitemShut {NoStop}%
\bibitem [{\citenamefont {Garrahan}\ and\ \citenamefont
  {Lesanovsky}(2010)}]{garrahan10a}%
  \BibitemOpen
  \bibfield  {author} {\bibinfo {author} {\bibfnamefont {J.~P.}\ \bibnamefont
  {Garrahan}}\ and\ \bibinfo {author} {\bibfnamefont {I.}~\bibnamefont
  {Lesanovsky}},\ }\bibfield  {title} {\enquote {\bibinfo {title}
  {Thermodynamics of quantum jump trajectories},}\ }\href
  {http://journals.aps.org/prl/abstract/10.1103/PhysRevLett.104.160601}
  {\bibfield  {journal} {\bibinfo  {journal} {Phys. Rev. Lett.}\ }\textbf
  {\bibinfo {volume} {104}},\ \bibinfo {pages} {160601} (\bibinfo {year}
  {2010})}\BibitemShut {NoStop}%
\bibitem [{\citenamefont {Ates}\ \emph {et~al.}(2012)\citenamefont {Ates},
  \citenamefont {Olmos}, \citenamefont {Garrahan},\ and\ \citenamefont
  {Lesanovsky}}]{ates12a}%
  \BibitemOpen
  \bibfield  {author} {\bibinfo {author} {\bibfnamefont {C.}~\bibnamefont
  {Ates}}, \bibinfo {author} {\bibfnamefont {B.}~\bibnamefont {Olmos}},
  \bibinfo {author} {\bibfnamefont {J.~P.}\ \bibnamefont {Garrahan}}, \ and\
  \bibinfo {author} {\bibfnamefont {I.}~\bibnamefont {Lesanovsky}},\ }\bibfield
   {title} {\enquote {\bibinfo {title} {Dynamical phases and intermittency of
  the dissipative quantum {Ising} model},}\ }\href
  {http://journals.aps.org/pra/abstract/10.1103/PhysRevA.85.043620} {\bibfield
  {journal} {\bibinfo  {journal} {Phys. Rev. A}\ }\textbf {\bibinfo {volume}
  {85}},\ \bibinfo {pages} {043620} (\bibinfo {year} {2012})}\BibitemShut
  {NoStop}%
\bibitem [{\citenamefont {Lesanovsky}\ \emph {et~al.}(2013)\citenamefont
  {Lesanovsky}, \citenamefont {van Horssen}, \citenamefont {Guta},\ and\
  \citenamefont {Garrahan}}]{lesanovsky13a}%
  \BibitemOpen
  \bibfield  {author} {\bibinfo {author} {\bibfnamefont {I.}~\bibnamefont
  {Lesanovsky}}, \bibinfo {author} {\bibfnamefont {M.}~\bibnamefont {van
  Horssen}}, \bibinfo {author} {\bibfnamefont {M.}~\bibnamefont {Guta}}, \ and\
  \bibinfo {author} {\bibfnamefont {J.~P.}\ \bibnamefont {Garrahan}},\
  }\bibfield  {title} {\enquote {\bibinfo {title} {Characterization of
  dynamical phase transitions in quantum jump trajectories beyond the
  properties of the stationary state},}\ }\href
  {http://journals.aps.org/prl/abstract/10.1103/PhysRevLett.110.150401}
  {\bibfield  {journal} {\bibinfo  {journal} {Phys. Rev. Lett.}\ }\textbf
  {\bibinfo {volume} {110}},\ \bibinfo {pages} {150401} (\bibinfo {year}
  {2013})}\BibitemShut {NoStop}%
\bibitem [{\citenamefont {Carollo}\ \emph {et~al.}(2017)\citenamefont
  {Carollo}, \citenamefont {Garrahan}, \citenamefont {Lesanovsky},\ and\
  \citenamefont {P{\'e}rez-Espigares}}]{carollo17a}%
  \BibitemOpen
  \bibfield  {author} {\bibinfo {author} {\bibfnamefont {F.}~\bibnamefont
  {Carollo}}, \bibinfo {author} {\bibfnamefont {J.P.}\ \bibnamefont
  {Garrahan}}, \bibinfo {author} {\bibfnamefont {I.}~\bibnamefont
  {Lesanovsky}}, \ and\ \bibinfo {author} {\bibfnamefont {C.}~\bibnamefont
  {P{\'e}rez-Espigares}},\ }\bibfield  {title} {\enquote {\bibinfo {title}
  {Fluctuating hydrodynamics, current fluctuations and hyperuniformity in
  boundary-driven open quantum chains},}\ }\href
  {https://arxiv.org/abs/1703.00355} {\bibfield  {journal} {\bibinfo  {journal}
  {arXiv:1703.00355}\ } (\bibinfo {year} {2017})}\BibitemShut {NoStop}%
\bibitem [{\citenamefont {Garrahan}\ \emph {et~al.}(2007)\citenamefont
  {Garrahan}, \citenamefont {Jack}, \citenamefont {Lecomte}, \citenamefont
  {Pitard}, \citenamefont {van Duijvendijk},\ and\ \citenamefont {van
  Wijland}}]{garrahan07a}%
  \BibitemOpen
  \bibfield  {author} {\bibinfo {author} {\bibfnamefont {J.~P.}\ \bibnamefont
  {Garrahan}}, \bibinfo {author} {\bibfnamefont {R.~L.}\ \bibnamefont {Jack}},
  \bibinfo {author} {\bibfnamefont {V.}~\bibnamefont {Lecomte}}, \bibinfo
  {author} {\bibfnamefont {E.}~\bibnamefont {Pitard}}, \bibinfo {author}
  {\bibfnamefont {K.}~\bibnamefont {van Duijvendijk}}, \ and\ \bibinfo {author}
  {\bibfnamefont {F.}~\bibnamefont {van Wijland}},\ }\bibfield  {title}
  {\enquote {\bibinfo {title} {Dynamical first-order phase transition in
  kinetically constrained models of glasses},}\ }\href
  {http://journals.aps.org/prl/abstract/10.1103/PhysRevLett.98.195702}
  {\bibfield  {journal} {\bibinfo  {journal} {Phys. Rev. Lett.}\ }\textbf
  {\bibinfo {volume} {98}},\ \bibinfo {pages} {195702} (\bibinfo {year}
  {2007})}\BibitemShut {NoStop}%
\bibitem [{\citenamefont {Garrahan}\ \emph {et~al.}(2009)\citenamefont
  {Garrahan}, \citenamefont {Jack}, \citenamefont {Lecomte}, \citenamefont
  {Pitard}, \citenamefont {van Duijvendijk},\ and\ \citenamefont {van
  Wijland}}]{garrahan09a}%
  \BibitemOpen
  \bibfield  {author} {\bibinfo {author} {\bibfnamefont {J.~P.}\ \bibnamefont
  {Garrahan}}, \bibinfo {author} {\bibfnamefont {R.~L.}\ \bibnamefont {Jack}},
  \bibinfo {author} {\bibfnamefont {V.}~\bibnamefont {Lecomte}}, \bibinfo
  {author} {\bibfnamefont {E.}~\bibnamefont {Pitard}}, \bibinfo {author}
  {\bibfnamefont {K.}~\bibnamefont {van Duijvendijk}}, \ and\ \bibinfo {author}
  {\bibfnamefont {F.}~\bibnamefont {van Wijland}},\ }\bibfield  {title}
  {\enquote {\bibinfo {title} {First-order dynamical phase transition in models
  of glasses: an approach based on ensembles of histories},}\ }\href
  {http://iopscience.iop.org/article/10.1088/1751-8113/42/7/075007} {\bibfield
  {journal} {\bibinfo  {journal} {J. Phys. A}\ }\textbf {\bibinfo {volume}
  {42}},\ \bibinfo {pages} {075007} (\bibinfo {year} {2009})}\BibitemShut
  {NoStop}%
\bibitem [{\citenamefont {Hedges}\ \emph {et~al.}(2009)\citenamefont {Hedges},
  \citenamefont {Jack}, \citenamefont {Garrahan},\ and\ \citenamefont
  {Chandler}}]{hedges09a}%
  \BibitemOpen
  \bibfield  {author} {\bibinfo {author} {\bibfnamefont {L.~O.}\ \bibnamefont
  {Hedges}}, \bibinfo {author} {\bibfnamefont {R.~L.}\ \bibnamefont {Jack}},
  \bibinfo {author} {\bibfnamefont {J.~P.}\ \bibnamefont {Garrahan}}, \ and\
  \bibinfo {author} {\bibfnamefont {D.}~\bibnamefont {Chandler}},\ }\bibfield
  {title} {\enquote {\bibinfo {title} {Dynamic order-disorder in atomistic
  models of structural glass formers},}\ }\href
  {http://science.sciencemag.org/content/323/5919/1309} {\bibfield  {journal}
  {\bibinfo  {journal} {Science}\ }\textbf {\bibinfo {volume} {323}},\ \bibinfo
  {pages} {1309} (\bibinfo {year} {2009})}\BibitemShut {NoStop}%
\bibitem [{\citenamefont {Chandler}\ and\ \citenamefont
  {Garrahan}(2010)}]{chandler10a}%
  \BibitemOpen
  \bibfield  {author} {\bibinfo {author} {\bibfnamefont {D.}~\bibnamefont
  {Chandler}}\ and\ \bibinfo {author} {\bibfnamefont {J.~P.}\ \bibnamefont
  {Garrahan}},\ }\bibfield  {title} {\enquote {\bibinfo {title} {Dynamics on
  the way to forming glass: bubbles in space-time.}}\ }\href
  {http://www.annualreviews.org/doi/abs/10.1146/annurev.physchem.040808.090405}
  {\bibfield  {journal} {\bibinfo  {journal} {Annu. Rev. Phys. Chem.}\ }\textbf
  {\bibinfo {volume} {61}},\ \bibinfo {pages} {191--217} (\bibinfo {year}
  {2010})}\BibitemShut {NoStop}%
\bibitem [{\citenamefont {Pitard}\ \emph {et~al.}(2011)\citenamefont {Pitard},
  \citenamefont {Lecomte},\ and\ \citenamefont {Van~Wijland}}]{pitard11a}%
  \BibitemOpen
  \bibfield  {author} {\bibinfo {author} {\bibfnamefont {E.}~\bibnamefont
  {Pitard}}, \bibinfo {author} {\bibfnamefont {V.}~\bibnamefont {Lecomte}}, \
  and\ \bibinfo {author} {\bibfnamefont {F.}~\bibnamefont {Van~Wijland}},\
  }\bibfield  {title} {\enquote {\bibinfo {title} {Dynamic transition in an
  atomic glass former: A molecular-dynamics evidence},}\ }\href
  {http://epljournal.edpsciences.org/articles/epl/abs/2011/23/epl14011/epl14011.html}
  {\bibfield  {journal} {\bibinfo  {journal} {Europhys. Lett.}\ }\textbf
  {\bibinfo {volume} {96}},\ \bibinfo {pages} {56002} (\bibinfo {year}
  {2011})}\BibitemShut {NoStop}%
\bibitem [{\citenamefont {Speck}\ \emph {et~al.}(2012)\citenamefont {Speck},
  \citenamefont {Malins},\ and\ \citenamefont {Royall}}]{speck12a}%
  \BibitemOpen
  \bibfield  {author} {\bibinfo {author} {\bibfnamefont {T.}~\bibnamefont
  {Speck}}, \bibinfo {author} {\bibfnamefont {A.}~\bibnamefont {Malins}}, \
  and\ \bibinfo {author} {\bibfnamefont {C.~P.}\ \bibnamefont {Royall}},\
  }\bibfield  {title} {\enquote {\bibinfo {title} {First-order phase transition
  in a model glass former: Coupling of local structure and dynamics},}\ }\href
  {http://journals.aps.org/prl/abstract/10.1103/PhysRevLett.109.195703}
  {\bibfield  {journal} {\bibinfo  {journal} {Phys. Rev. Lett.}\ }\textbf
  {\bibinfo {volume} {109}},\ \bibinfo {pages} {195703} (\bibinfo {year}
  {2012})}\BibitemShut {NoStop}%
\bibitem [{\citenamefont {Pinchaipat}\ \emph {et~al.}(2017)\citenamefont
  {Pinchaipat}, \citenamefont {Campo}, \citenamefont {Turci}, \citenamefont
  {Hallett}, \citenamefont {Speck},\ and\ \citenamefont
  {Royall}}]{pinchaipat17a}%
  \BibitemOpen
  \bibfield  {author} {\bibinfo {author} {\bibfnamefont {R.}~\bibnamefont
  {Pinchaipat}}, \bibinfo {author} {\bibfnamefont {M.}~\bibnamefont {Campo}},
  \bibinfo {author} {\bibfnamefont {F.}~\bibnamefont {Turci}}, \bibinfo
  {author} {\bibfnamefont {J.}~\bibnamefont {Hallett}}, \bibinfo {author}
  {\bibfnamefont {T.}~\bibnamefont {Speck}}, \ and\ \bibinfo {author}
  {\bibfnamefont {C.~P.}\ \bibnamefont {Royall}},\ }\bibfield  {title}
  {\enquote {\bibinfo {title} {Experimental evidence for a structural-dynamical
  transition in trajectory space},}\ }\href
  {https://journals.aps.org/prl/abstract/10.1103/PhysRevLett.119.028004}
  {\bibfield  {journal} {\bibinfo  {journal} {Phys. Rev. Lett.}\ }\textbf
  {\bibinfo {volume} {119}},\ \bibinfo {pages} {028004} (\bibinfo {year}
  {2017})}\BibitemShut {NoStop}%
\bibitem [{\citenamefont {Abou}\ \emph {et~al.}(2017)\citenamefont {Abou},
  \citenamefont {Colin}, \citenamefont {Lecomte}, \citenamefont {Pitard},\ and\
  \citenamefont {van Wijland}}]{abou17a}%
  \BibitemOpen
  \bibfield  {author} {\bibinfo {author} {\bibfnamefont {B.}~\bibnamefont
  {Abou}}, \bibinfo {author} {\bibfnamefont {R.}~\bibnamefont {Colin}},
  \bibinfo {author} {\bibfnamefont {V.}~\bibnamefont {Lecomte}}, \bibinfo
  {author} {\bibfnamefont {E.}~\bibnamefont {Pitard}}, \ and\ \bibinfo {author}
  {\bibfnamefont {F.}~\bibnamefont {van Wijland}},\ }\bibfield  {title}
  {\enquote {\bibinfo {title} {Activity statistics in a colloidal glass former:
  experimental evidence for a dynamical transition},}\ }\href
  {https://arxiv.org/abs/1705.00855} {\bibfield  {journal} {\bibinfo  {journal}
  {arXiv:1705.00855}\ } (\bibinfo {year} {2017})}\BibitemShut {NoStop}%
\bibitem [{\citenamefont {Garrahan}\ \emph {et~al.}(2011)\citenamefont
  {Garrahan}, \citenamefont {Armour},\ and\ \citenamefont
  {Lesanovsky}}]{garrahan11a}%
  \BibitemOpen
  \bibfield  {author} {\bibinfo {author} {\bibfnamefont {J.~P.}\ \bibnamefont
  {Garrahan}}, \bibinfo {author} {\bibfnamefont {A.~D.}\ \bibnamefont
  {Armour}}, \ and\ \bibinfo {author} {\bibfnamefont {I.}~\bibnamefont
  {Lesanovsky}},\ }\bibfield  {title} {\enquote {\bibinfo {title} {Quantum
  trajectory phase transitions in the micromaser},}\ }\href
  {http://journals.aps.org/pre/abstract/10.1103/PhysRevE.84.021115} {\bibfield
  {journal} {\bibinfo  {journal} {Phys. Rev. E}\ }\textbf {\bibinfo {volume}
  {84}},\ \bibinfo {pages} {021115} (\bibinfo {year} {2011})}\BibitemShut
  {NoStop}%
\bibitem [{\citenamefont {Genway}\ \emph {et~al.}(2012)\citenamefont {Genway},
  \citenamefont {Garrahan}, \citenamefont {Lesanovsky},\ and\ \citenamefont
  {Armour}}]{genway12a}%
  \BibitemOpen
  \bibfield  {author} {\bibinfo {author} {\bibfnamefont {S.}~\bibnamefont
  {Genway}}, \bibinfo {author} {\bibfnamefont {J.~P.}\ \bibnamefont
  {Garrahan}}, \bibinfo {author} {\bibfnamefont {I.}~\bibnamefont
  {Lesanovsky}}, \ and\ \bibinfo {author} {\bibfnamefont {A.~D.}\ \bibnamefont
  {Armour}},\ }\bibfield  {title} {\enquote {\bibinfo {title} {Phase
  transitions in trajectories of a superconducting single-electron transistor
  coupled to a resonator},}\ }\href
  {http://journals.aps.org/pre/abstract/10.1103/PhysRevE.85.051122} {\bibfield
  {journal} {\bibinfo  {journal} {Phys. Rev. E}\ }\textbf {\bibinfo {volume}
  {85}},\ \bibinfo {pages} {051122} (\bibinfo {year} {2012})}\BibitemShut
  {NoStop}%
\bibitem [{\citenamefont {Manzano}\ and\ \citenamefont
  {Hurtado}(2014)}]{manzano14a}%
  \BibitemOpen
  \bibfield  {author} {\bibinfo {author} {\bibfnamefont {D.}~\bibnamefont
  {Manzano}}\ and\ \bibinfo {author} {\bibfnamefont {P.~I.}\ \bibnamefont
  {Hurtado}},\ }\bibfield  {title} {\enquote {\bibinfo {title} {Symmetry and
  the thermodynamics of currents in open quantum systems},}\ }\href
  {https://journals.aps.org/prb/abstract/10.1103/PhysRevB.90.125138} {\bibfield
   {journal} {\bibinfo  {journal} {Phys. Rev. B}\ }\textbf {\bibinfo {volume}
  {90}},\ \bibinfo {pages} {125138} (\bibinfo {year} {2014})}\BibitemShut
  {NoStop}%
\bibitem [{\citenamefont {Manzano}\ and\ \citenamefont
  {Kyoseva}(2016)}]{manzano16a}%
  \BibitemOpen
  \bibfield  {author} {\bibinfo {author} {\bibfnamefont {D.}~\bibnamefont
  {Manzano}}\ and\ \bibinfo {author} {\bibfnamefont {E.}~\bibnamefont
  {Kyoseva}},\ }\bibfield  {title} {\enquote {\bibinfo {title} {An atomic
  symmetry-controlled thermal switch},}\ }\href
  {https://www.nature.com/articles/srep31161} {\bibfield  {journal} {\bibinfo
  {journal} {Sci. Rep.}\ }\textbf {\bibinfo {volume} {6}},\ \bibinfo {pages}
  {31161} (\bibinfo {year} {2016})}\BibitemShut {NoStop}%
\bibitem [{\citenamefont {Manzano}\ and\ \citenamefont
  {Hurtado}()}]{manzano17a}%
  \BibitemOpen
  \bibfield  {author} {\bibinfo {author} {\bibfnamefont {D.}~\bibnamefont
  {Manzano}}\ and\ \bibinfo {author} {\bibfnamefont {P.I.}\ \bibnamefont
  {Hurtado}},\ }\bibfield  {title} {\enquote {\bibinfo {title} {Harnessing
  symmetry to control quantum transport},}\ }\href
  {https://arxiv.org/abs/1706.xxxxx} {\bibinfo  {journal} {arXiv:1706.xxxxx}\
  }\BibitemShut {NoStop}%
\bibitem [{\citenamefont {Touchette}(2009)}]{touchette09a}%
  \BibitemOpen
\bibfield  {journal} {  }\bibfield  {author} {\bibinfo {author} {\bibfnamefont
  {H.}~\bibnamefont {Touchette}},\ }\bibfield  {title} {\enquote {\bibinfo
  {title} {The large deviation approach to statistical mechanics},}\ }\href
  {http://dx.doi.org/10.1016/j.physrep.2009.05.002} {\bibfield  {journal}
  {\bibinfo  {journal} {Phys. Rep.}\ }\textbf {\bibinfo {volume} {478}},\
  \bibinfo {pages} {1--69} (\bibinfo {year} {2009})}\BibitemShut {NoStop}%
\bibitem [{\citenamefont {Yang}\ and\ \citenamefont {Lee}(1952)}]{yang52a}%
  \BibitemOpen
  \bibfield  {author} {\bibinfo {author} {\bibfnamefont {C.-N.}\ \bibnamefont
  {Yang}}\ and\ \bibinfo {author} {\bibfnamefont {T.-D.}\ \bibnamefont {Lee}},\
  }\bibfield  {title} {\enquote {\bibinfo {title} {Statistical theory of
  equations of state and phase transitions. i. {T}heory of condensation},}\
  }\href {https://journals.aps.org/pr/abstract/10.1103/PhysRev.87.404}
  {\bibfield  {journal} {\bibinfo  {journal} {Phys. Rev.}\ }\textbf {\bibinfo
  {volume} {87}},\ \bibinfo {pages} {404} (\bibinfo {year} {1952})}\BibitemShut
  {NoStop}%
\bibitem [{\citenamefont {Arndt}(2000)}]{arndt00a}%
  \BibitemOpen
  \bibfield  {author} {\bibinfo {author} {\bibfnamefont {P.F.}\ \bibnamefont
  {Arndt}},\ }\bibfield  {title} {\enquote {\bibinfo {title} {{Y}ang-{L}ee
  theory for a nonequilibrium phase transition},}\ }\href
  {https://journals.aps.org/prl/abstract/10.1103/PhysRevLett.84.814} {\bibfield
   {journal} {\bibinfo  {journal} {Phys. Rev. Lett.}\ }\textbf {\bibinfo
  {volume} {84}},\ \bibinfo {pages} {814} (\bibinfo {year} {2000})}\BibitemShut
  {NoStop}%
\bibitem [{\citenamefont {Blythe}\ and\ \citenamefont
  {Evans}(2002)}]{blythe02a}%
  \BibitemOpen
  \bibfield  {author} {\bibinfo {author} {\bibfnamefont {R.~A.}\ \bibnamefont
  {Blythe}}\ and\ \bibinfo {author} {\bibfnamefont {M.~R.}\ \bibnamefont
  {Evans}},\ }\bibfield  {title} {\enquote {\bibinfo {title} {{L}ee-{Y}ang
  zeros and phase transitions in nonequilibrium steady states},}\ }\href
  {https://journals.aps.org/prl/abstract/10.1103/PhysRevLett.89.080601}
  {\bibfield  {journal} {\bibinfo  {journal} {Phys. Rev. Lett.}\ }\textbf
  {\bibinfo {volume} {89}},\ \bibinfo {pages} {080601} (\bibinfo {year}
  {2002})}\BibitemShut {NoStop}%
\bibitem [{\citenamefont {Dammer}\ \emph {et~al.}(2002)\citenamefont {Dammer},
  \citenamefont {Dahmen},\ and\ \citenamefont {Hinrichsen}}]{dammer02a}%
  \BibitemOpen
  \bibfield  {author} {\bibinfo {author} {\bibfnamefont {S.M.}\ \bibnamefont
  {Dammer}}, \bibinfo {author} {\bibfnamefont {S.R.}\ \bibnamefont {Dahmen}}, \
  and\ \bibinfo {author} {\bibfnamefont {H.}~\bibnamefont {Hinrichsen}},\
  }\bibfield  {title} {\enquote {\bibinfo {title} {{Y}ang-{L}ee zeros for a
  nonequilibrium phase transition},}\ }\href
  {http://iopscience.iop.org/article/10.1088/0305-4470/35/21/303/meta}
  {\bibfield  {journal} {\bibinfo  {journal} {J. Phys. A}\ }\textbf {\bibinfo
  {volume} {35}},\ \bibinfo {pages} {4527} (\bibinfo {year}
  {2002})}\BibitemShut {NoStop}%
\bibitem [{\citenamefont {Blythe}\ and\ \citenamefont
  {Evans}(2003)}]{blythe03a}%
  \BibitemOpen
  \bibfield  {author} {\bibinfo {author} {\bibfnamefont {R.A.}\ \bibnamefont
  {Blythe}}\ and\ \bibinfo {author} {\bibfnamefont {M.R.}\ \bibnamefont
  {Evans}},\ }\bibfield  {title} {\enquote {\bibinfo {title} {The {L}ee-{Y}ang
  theory of equilibrium and nonequilibrium phase transitions},}\ }\href
  {http://www.scielo.br/scielo.php?script=sci_arttext&pid=S0103-97332003000300008&lng=en&nrm=iso&tlng=en}
  {\bibfield  {journal} {\bibinfo  {journal} {Braz. J. Phys.}\ }\textbf
  {\bibinfo {volume} {33}},\ \bibinfo {pages} {464} (\bibinfo {year}
  {2003})}\BibitemShut {NoStop}%
\bibitem [{\citenamefont {Flindt}\ and\ \citenamefont
  {Garrahan}(2013)}]{flindt13a}%
  \BibitemOpen
  \bibfield  {author} {\bibinfo {author} {\bibfnamefont {C.}~\bibnamefont
  {Flindt}}\ and\ \bibinfo {author} {\bibfnamefont {J.P.}\ \bibnamefont
  {Garrahan}},\ }\bibfield  {title} {\enquote {\bibinfo {title} {Trajectory
  phase transitions, {L}ee-{Y}ang zeros, and high-order cumulants in full
  counting statistics},}\ }\href
  {https://journals.aps.org/prl/abstract/10.1103/PhysRevLett.110.050601}
  {\bibfield  {journal} {\bibinfo  {journal} {Phys. Rev. Lett.}\ }\textbf
  {\bibinfo {volume} {110}},\ \bibinfo {pages} {050601} (\bibinfo {year}
  {2013})}\BibitemShut {NoStop}%
\bibitem [{\citenamefont {Hickey}\ \emph {et~al.}(2014)\citenamefont {Hickey},
  \citenamefont {Flindt},\ and\ \citenamefont {Garrahan}}]{hickey14a}%
  \BibitemOpen
  \bibfield  {author} {\bibinfo {author} {\bibfnamefont {J.M.}\ \bibnamefont
  {Hickey}}, \bibinfo {author} {\bibfnamefont {C.}~\bibnamefont {Flindt}}, \
  and\ \bibinfo {author} {\bibfnamefont {J.P.}\ \bibnamefont {Garrahan}},\
  }\bibfield  {title} {\enquote {\bibinfo {title} {Intermittency and dynamical
  {L}ee-{Y}ang zeros of open quantum systems},}\ }\href
  {https://journals.aps.org/pre/abstract/10.1103/PhysRevE.90.062128} {\bibfield
   {journal} {\bibinfo  {journal} {Phys. Rev. E}\ }\textbf {\bibinfo {volume}
  {90}} (\bibinfo {year} {2014})}\BibitemShut {NoStop}%
\bibitem [{\citenamefont {Brandner}\ \emph {et~al.}(2017)\citenamefont
  {Brandner}, \citenamefont {Maisi}, \citenamefont {Pekola}, \citenamefont
  {Garrahan},\ and\ \citenamefont {Flindt}}]{brandner17a}%
  \BibitemOpen
  \bibfield  {author} {\bibinfo {author} {\bibfnamefont {K.}~\bibnamefont
  {Brandner}}, \bibinfo {author} {\bibfnamefont {V.F.}\ \bibnamefont {Maisi}},
  \bibinfo {author} {\bibfnamefont {J.P.}\ \bibnamefont {Pekola}}, \bibinfo
  {author} {\bibfnamefont {J.P.}\ \bibnamefont {Garrahan}}, \ and\ \bibinfo
  {author} {\bibfnamefont {C.}~\bibnamefont {Flindt}},\ }\bibfield  {title}
  {\enquote {\bibinfo {title} {Experimental determination of dynamical
  {L}ee-{Y}ang zeros},}\ }\href
  {https://journals.aps.org/prl/abstract/10.1103/PhysRevLett.118.180601}
  {\bibfield  {journal} {\bibinfo  {journal} {Phys. Rev. Lett.}\ }\textbf
  {\bibinfo {volume} {118}} (\bibinfo {year} {2017})}\BibitemShut {NoStop}%
\bibitem [{\citenamefont {Barr{\'e}}\ \emph {et~al.}(2017)\citenamefont
  {Barr{\'e}}, \citenamefont {Bernardin},\ and\ \citenamefont
  {Chetrite}}]{barre17a}%
  \BibitemOpen
  \bibfield  {author} {\bibinfo {author} {\bibfnamefont {J.}~\bibnamefont
  {Barr{\'e}}}, \bibinfo {author} {\bibfnamefont {C.}~\bibnamefont
  {Bernardin}}, \ and\ \bibinfo {author} {\bibfnamefont {R.}~\bibnamefont
  {Chetrite}},\ }\bibfield  {title} {\enquote {\bibinfo {title} {Density large
  deviations for multidimensional stochastic hyperbolic conservation laws},}\
  }\href {https://arxiv.org/abs/1702.03769} {\bibfield  {journal} {\bibinfo
  {journal} {arXiv:1702.03769}\ } (\bibinfo {year} {2017})}\BibitemShut
  {NoStop}%
\bibitem [{\citenamefont {Lam}\ \emph {et~al.}(2009)\citenamefont {Lam},
  \citenamefont {Kurchan},\ and\ \citenamefont {Levine}}]{lam09a}%
  \BibitemOpen
  \bibfield  {author} {\bibinfo {author} {\bibfnamefont {K.~D. N.~T.}\
  \bibnamefont {Lam}}, \bibinfo {author} {\bibfnamefont {J.}~\bibnamefont
  {Kurchan}}, \ and\ \bibinfo {author} {\bibfnamefont {D.}~\bibnamefont
  {Levine}},\ }\bibfield  {title} {\enquote {\bibinfo {title} {Order in
  extremal trajectories},}\ }\href
  {http://link.springer.com/article/10.1007/s10955-009-9828-8} {\bibfield
  {journal} {\bibinfo  {journal} {J. Stat. Phys.}\ }\textbf {\bibinfo {volume}
  {137}},\ \bibinfo {pages} {1079--1093} (\bibinfo {year} {2009})}\BibitemShut
  {NoStop}%
\bibitem [{\citenamefont {Chetrite}\ and\ \citenamefont
  {Touchette}(2015{\natexlab{a}})}]{chetrite15a}%
  \BibitemOpen
  \bibfield  {author} {\bibinfo {author} {\bibfnamefont {R.}~\bibnamefont
  {Chetrite}}\ and\ \bibinfo {author} {\bibfnamefont {H.}~\bibnamefont
  {Touchette}},\ }\bibfield  {title} {\enquote {\bibinfo {title} {Variational
  and optimal control representations of conditioned and driven processes},}\
  }\href
  {http://iopscience.iop.org/article/10.1088/1742-5468/2015/12/P12001/meta}
  {\bibfield  {journal} {\bibinfo  {journal} {J. Stat. Mech. P12001}\ }
  (\bibinfo {year} {2015}{\natexlab{a}})}\BibitemShut {NoStop}%
\bibitem [{\citenamefont {Chetrite}\ and\ \citenamefont
  {Touchette}(2015{\natexlab{b}})}]{chetrite15b}%
  \BibitemOpen
  \bibfield  {author} {\bibinfo {author} {\bibfnamefont {R.}~\bibnamefont
  {Chetrite}}\ and\ \bibinfo {author} {\bibfnamefont {H.}~\bibnamefont
  {Touchette}},\ }\bibfield  {title} {\enquote {\bibinfo {title}
  {Nonequilibrium {Markov} processes conditioned on large deviations},}\ }\href
  {https://link.springer.com/article/10.1007%2Fs00023-014-0375-8} {\bibfield
  {journal} {\bibinfo  {journal} {Ann. Henri Poincare}\ }\textbf {\bibinfo
  {volume} {16}},\ \bibinfo {pages} {2005} (\bibinfo {year}
  {2015}{\natexlab{b}})}\BibitemShut {NoStop}%
\bibitem [{\citenamefont {Derrida}(1998)}]{derrida98a}%
  \BibitemOpen
  \bibfield  {author} {\bibinfo {author} {\bibfnamefont {B.}~\bibnamefont
  {Derrida}},\ }\bibfield  {title} {\enquote {\bibinfo {title} {An exactly
  soluble non-equilibrium system: {The} asymmetric simple exclusion process},}\
  }\href {http://www.sciencedirect.com/science/article/pii/S0370157398000064}
  {\bibfield  {journal} {\bibinfo  {journal} {Phys. Rep.}\ }\textbf {\bibinfo
  {volume} {301}},\ \bibinfo {pages} {65--83} (\bibinfo {year}
  {1998})}\BibitemShut {NoStop}%
\bibitem [{\citenamefont {Giardin\`a}\ \emph {et~al.}(2006)\citenamefont
  {Giardin\`a}, \citenamefont {Kurchan},\ and\ \citenamefont
  {Peliti}}]{giardina06a}%
  \BibitemOpen
  \bibfield  {author} {\bibinfo {author} {\bibfnamefont {C.}~\bibnamefont
  {Giardin\`a}}, \bibinfo {author} {\bibfnamefont {J.}~\bibnamefont {Kurchan}},
  \ and\ \bibinfo {author} {\bibfnamefont {L.}~\bibnamefont {Peliti}},\
  }\bibfield  {title} {\enquote {\bibinfo {title} {Direct evaluation of
  large-deviation functions},}\ }\href
  {http://journals.aps.org/prl/abstract/10.1103/PhysRevLett.96.120603}
  {\bibfield  {journal} {\bibinfo  {journal} {Phys. Rev. Lett.}\ }\textbf
  {\bibinfo {volume} {96}},\ \bibinfo {pages} {120603} (\bibinfo {year}
  {2006})}\BibitemShut {NoStop}%
\bibitem [{\citenamefont {Lecomte}\ and\ \citenamefont
  {Tailleur}()}]{lecomte07a}%
  \BibitemOpen
  \bibfield  {author} {\bibinfo {author} {\bibfnamefont {V.}~\bibnamefont
  {Lecomte}}\ and\ \bibinfo {author} {\bibfnamefont {J.}~\bibnamefont
  {Tailleur}},\ }\bibfield  {title} {\enquote {\bibinfo {title} {A numerical
  approach to large deviations in continuous time},}\ }\href
  {http://iopscience.iop.org/article/10.1088/1742-5468/2007/03/P03004/meta}
  {\bibinfo  {journal} {J. Stat. Mech. P03004 (2007)}\ }\BibitemShut {NoStop}%
\bibitem [{\citenamefont {Giardin\`a}\ \emph {et~al.}(2011)\citenamefont
  {Giardin\`a}, \citenamefont {Kurchan}, \citenamefont {Lecomte},\ and\
  \citenamefont {Tailleur}}]{giardina11a}%
  \BibitemOpen
\bibfield  {journal} {  }\bibfield  {author} {\bibinfo {author} {\bibfnamefont
  {C.}~\bibnamefont {Giardin\`a}}, \bibinfo {author} {\bibfnamefont
  {J.}~\bibnamefont {Kurchan}}, \bibinfo {author} {\bibfnamefont
  {V.}~\bibnamefont {Lecomte}}, \ and\ \bibinfo {author} {\bibfnamefont
  {J.}~\bibnamefont {Tailleur}},\ }\bibfield  {title} {\enquote {\bibinfo
  {title} {Simulating rare events in dynamical processes},}\ }\href
  {http://link.springer.com/article/10.1007%2Fs10955-011-0350-4} {\bibfield
  {journal} {\bibinfo  {journal} {J. Stat. Phys.}\ }\textbf {\bibinfo {volume}
  {145}},\ \bibinfo {pages} {787--811} (\bibinfo {year} {2011})}\BibitemShut
  {NoStop}%
\bibitem [{Note1()}]{Note1}%
  \BibitemOpen
  \bibinfo {note} {The rates $r^\alpha _{\pm }\equiv \protect \REV@text
  {exp}[\pm E_\alpha /L]/2$ converge for large $L$ to the standard ones found
  in literature \cite {bodineau05a,derrida07a}, namely ${\begingroup 1\endgroup
  \over 2}(1 \pm E_\alpha /L)$, but avoid problems with negative rates for
  small $L$. Indeed, the hydrodynamic description of both variants of the model
  is identical in the thermodynamic limit. However, for finite, moderate values
  of $L$ the field per unit length (${\protect \bf {E}}/L$) is too strong,
  leading to an \protect \emph {effective anisotropy} in the system. In fact,
  by expanding the microscopic transition rate $r_\pm ^\alpha $ to second order
  in the field per unit length, i.e. $r_\pm ^\alpha \approx {\begingroup
  1\endgroup \over 2}[1\pm E_\alpha /L+{\begingroup 1\endgroup \over
  2}(E_\alpha /L)^2] + {\protect \cal O}[(E_\alpha /L)^3]$, it is easy to show
  using a simple random walk argument that the second-order perturbation
  results in an effective increase of diffusivity and mobility along the field
  direction, and an associated decrease in the orthogonal
  direction.}\BibitemShut {Stop}%
\bibitem [{SMp()}]{SMprl}%
  \BibitemOpen
  \href@noop {} {\bibinfo  {journal} {See Supplementary Material
  http://link.aps.org/ supplemental/XXXX/PhysRevLett.XXX for the details}\
  }\BibitemShut {NoStop}%
\bibitem [{\citenamefont {P\'erez-Espigares}\ \emph {et~al.}(2016)\citenamefont
  {P\'erez-Espigares}, \citenamefont {Garrido},\ and\ \citenamefont
  {Hurtado}}]{perez-espigares16a}%
  \BibitemOpen
\bibfield  {journal} {  }\bibfield  {author} {\bibinfo {author} {\bibfnamefont
  {C.}~\bibnamefont {P\'erez-Espigares}}, \bibinfo {author} {\bibfnamefont
  {P.~L.}\ \bibnamefont {Garrido}}, \ and\ \bibinfo {author} {\bibfnamefont
  {P.~I.}\ \bibnamefont {Hurtado}},\ }\bibfield  {title} {\enquote {\bibinfo
  {title} {Weak additivity principle for current statistics in
  $d$-dimensions},}\ }\href
  {http://journals.aps.org/pre/abstract/10.1103/PhysRevE.93.040103} {\bibfield
  {journal} {\bibinfo  {journal} {Phys. Rev. E}\ }\textbf {\bibinfo {volume}
  {93}},\ \bibinfo {pages} {040103(R)} (\bibinfo {year} {2016})}\BibitemShut
  {NoStop}%
\bibitem [{\citenamefont {Villavicencio-Sanchez}\ and\ \citenamefont
  {Harris}(2016)}]{villavicencio16a}%
  \BibitemOpen
  \bibfield  {author} {\bibinfo {author} {\bibfnamefont {R.}~\bibnamefont
  {Villavicencio-Sanchez}}\ and\ \bibinfo {author} {\bibfnamefont {R.~J.}\
  \bibnamefont {Harris}},\ }\bibfield  {title} {\enquote {\bibinfo {title}
  {Local structure of current fluctuations in diffusive systems beyond one
  dimension},}\ }\href
  {http://journals.aps.org/pre/abstract/10.1103/PhysRevE.93.032134} {\bibfield
  {journal} {\bibinfo  {journal} {Phys. Rev. E}\ }\textbf {\bibinfo {volume}
  {93}},\ \bibinfo {pages} {032134} (\bibinfo {year} {2016})}\BibitemShut
  {NoStop}%
\bibitem [{\citenamefont {Tiz{\'o}n-Escamilla}\ \emph
  {et~al.}(2017)\citenamefont {Tiz{\'o}n-Escamilla}, \citenamefont {Hurtado},\
  and\ \citenamefont {Garrido}}]{tizon-escamilla17a}%
  \BibitemOpen
  \bibfield  {author} {\bibinfo {author} {\bibfnamefont {N.}~\bibnamefont
  {Tiz{\'o}n-Escamilla}}, \bibinfo {author} {\bibfnamefont {P.~I.}\
  \bibnamefont {Hurtado}}, \ and\ \bibinfo {author} {\bibfnamefont {P.~L.}\
  \bibnamefont {Garrido}},\ }\bibfield  {title} {\enquote {\bibinfo {title}
  {Structure of the optimal path to a fluctuation},}\ }\href
  {http://journals.aps.org/pre/pdf/10.1103/PhysRevE.95.002100} {\bibfield
  {journal} {\bibinfo  {journal} {Phys. Rev. E}\ }\textbf {\bibinfo {volume}
  {95}},\ \bibinfo {pages} {002100} (\bibinfo {year} {2017})}\BibitemShut
  {NoStop}%
\bibitem [{\citenamefont {Hurtado}\ and\ \citenamefont
  {Garrido}(2009)}]{hurtado09a}%
  \BibitemOpen
  \bibfield  {author} {\bibinfo {author} {\bibfnamefont {P.~I.}\ \bibnamefont
  {Hurtado}}\ and\ \bibinfo {author} {\bibfnamefont {P.~L.}\ \bibnamefont
  {Garrido}},\ }\bibfield  {title} {\enquote {\bibinfo {title} {Current
  fluctuations and statistics during a large deviation event in an exactly
  solvable transport model},}\ }\href
  {http://iopscience.iop.org/article/10.1088/1742-5468/2009/02/P02032/meta}
  {\bibfield  {journal} {\bibinfo  {journal} {J. Stat. Mech. P02032}\ }
  (\bibinfo {year} {2009})}\BibitemShut {NoStop}%
\bibitem [{\citenamefont {Nemoto}\ \emph {et~al.}(2016)\citenamefont {Nemoto},
  \citenamefont {Bouchet}, \citenamefont {Jack},\ and\ \citenamefont
  {Lecomte}}]{nemoto16a}%
  \BibitemOpen
  \bibfield  {author} {\bibinfo {author} {\bibfnamefont {T.}~\bibnamefont
  {Nemoto}}, \bibinfo {author} {\bibfnamefont {F.}~\bibnamefont {Bouchet}},
  \bibinfo {author} {\bibfnamefont {R.~L.}\ \bibnamefont {Jack}}, \ and\
  \bibinfo {author} {\bibfnamefont {V.}~\bibnamefont {Lecomte}},\ }\bibfield
  {title} {\enquote {\bibinfo {title} {Population dynamics method with a
  multi-canonical feedback control},}\ }\href
  {http://journals.aps.org/pre/abstract/10.1103/PhysRevE.93.062123} {\bibfield
  {journal} {\bibinfo  {journal} {Phys. Rev. E}\ }\textbf {\bibinfo {volume}
  {93}},\ \bibinfo {pages} {062123} (\bibinfo {year} {2016})}\BibitemShut
  {NoStop}%
\bibitem [{\citenamefont {Felderhof}(1971)}]{felderhof71a}%
  \BibitemOpen
  \bibfield  {author} {\bibinfo {author} {\bibfnamefont {B.U.}\ \bibnamefont
  {Felderhof}},\ }\bibfield  {title} {\enquote {\bibinfo {title} {Spin
  relaxation of the {Ising} chain},}\ }\href
  {http://www.sciencedirect.com/science/article/pii/S003448777180006X}
  {\bibfield  {journal} {\bibinfo  {journal} {Rep. Math. Phys.}\ }\textbf
  {\bibinfo {volume} {1}},\ \bibinfo {pages} {215} (\bibinfo {year}
  {1971})}\BibitemShut {NoStop}%
\bibitem [{\citenamefont {Alcaraz}\ \emph {et~al.}(1994)\citenamefont
  {Alcaraz}, \citenamefont {Droz}, \citenamefont {Henkel},\ and\ \citenamefont
  {Rittenberg}}]{alcaraz94a}%
  \BibitemOpen
  \bibfield  {author} {\bibinfo {author} {\bibfnamefont {F.C.}\ \bibnamefont
  {Alcaraz}}, \bibinfo {author} {\bibfnamefont {M.}~\bibnamefont {Droz}},
  \bibinfo {author} {\bibfnamefont {M.}~\bibnamefont {Henkel}}, \ and\ \bibinfo
  {author} {\bibfnamefont {V.}~\bibnamefont {Rittenberg}},\ }\bibfield  {title}
  {\enquote {\bibinfo {title} {Reaction-diffusion processes, critical dynamics
  and quantum chains},}\ }\href
  {http://www.sciencedirect.com/science/article/pii/S0003491684710268?via%3Dihub}
  {\bibfield  {journal} {\bibinfo  {journal} {Ann. Phys.}\ }\textbf {\bibinfo
  {volume} {230}},\ \bibinfo {pages} {250} (\bibinfo {year}
  {1994})}\BibitemShut {NoStop}%
\bibitem [{\citenamefont {Stinchcombe}\ and\ \citenamefont
  {Sch{\"u}tz}(1995)}]{stinchcombe95a}%
  \BibitemOpen
  \bibfield  {author} {\bibinfo {author} {\bibfnamefont {R.B.}\ \bibnamefont
  {Stinchcombe}}\ and\ \bibinfo {author} {\bibfnamefont {G.M.}\ \bibnamefont
  {Sch{\"u}tz}},\ }\bibfield  {title} {\enquote {\bibinfo {title} {Application
  of operator algebras to stochastic dynamics and the {H}eisenberg chain},}\
  }\href {http://journals.aps.org/prl/abstract/10.1103/PhysRevLett.75.140}
  {\bibfield  {journal} {\bibinfo  {journal} {Phys. Rev. Lett.}\ }\textbf
  {\bibinfo {volume} {75}},\ \bibinfo {pages} {140} (\bibinfo {year}
  {1995})}\BibitemShut {NoStop}%
\bibitem [{\citenamefont {Stinchcombe}(2001)}]{stinchcombe01a}%
  \BibitemOpen
  \bibfield  {author} {\bibinfo {author} {\bibfnamefont {R.B.}\ \bibnamefont
  {Stinchcombe}},\ }\bibfield  {title} {\enquote {\bibinfo {title} {Stochastic
  non-equilibrium systems},}\ }\href
  {http://www.tandfonline.com/doi/abs/10.1080/00018730110099650} {\bibfield
  {journal} {\bibinfo  {journal} {Adv. Phys.}\ }\textbf {\bibinfo {volume}
  {50}},\ \bibinfo {pages} {431} (\bibinfo {year} {2001})}\BibitemShut
  {NoStop}%
\end{thebibliography}

\begin{thebibliography}{10}

\bibitem{Bertini1} L. Bertini, A. De Sole, D. Gabrielli, G. Jona-Lasinio and C. Landim, ``Macroscopic fluctuation theory," \href{http://journals.aps.org/rmp/abstract/10.1103/RevModPhys.87.593}{Rev. Mod. Phys. {\bf 87}, 593 (2015)}.

\bibitem{Derrida1} B. Derrida, ``Non-equilibrium steady states: fluctuations and large deviations of the density and of the current," \href{http://iopscience.iop.org/article/10.1088/1742-5468/2007/07/P07023}{J. Stat. Mech. P07023 (2007)}.

\bibitem{weJSP} P. I. Hurtado, C. P\'erez-Espigares, J. J. del Pozo, and P. L. Garrido, ``Thermodynamics of currents in nonequilibrium diffusive systems: theory and simulation," \href{http://link.springer.com/article/10.1007/s10955-013-0894-6}{J. Stat. Phys. {\bf 154}, 214 (2014)}.

\bibitem{weIFR} P. I. Hurtado, C. P\'erez-Espigares, J. J. del Pozo, and P. L. Garrido, ``Symmetries in fluctuations far from equilibrium," \href{http://www.pnas.org/content/108/19/7704.short}{Proc. Natl. Acad. Sci. USA {\bf 108}, 7704 (2011)}.

\bibitem{AFR} R. Villavicencio-S\'anchez, R. J. Harris, and H. Touchette, ``Fluctuation relations for anisotropic systems," \href{http://iopscience.iop.org/article/10.1209/0295-5075/105/30009/meta}{Europhys. Lett. {\bf 105}, 30009 (2014)}.

\bibitem{SFT} C. P\'erez-Espigares, F. Redig, and C. Giardin\`a, ``Spatial fluctuation theorem," \href{http://iopscience.iop.org/article/10.1088/1751-8113/48/35/35FT01/meta}{J. Phys. A {\bf 48}, 35FT01 (2015)}.

\bibitem{kmp} C. Kipnis, C. Marchioro and E. Presutti, ``Heat flow in an exactly solvable model," \href{http://link.springer.com/article/10.1007/BF01011740}{J. Stat. Phys. {\bf 27}, 65 (1982)}.

\bibitem{wasep} B. Derrida, ``An exactly soluble non-equilibrium system: The asymmetric simple exclusion process,Ó \href{http://www.sciencedirect.com/science/article/pii/S0370157398000064}{Phys. Rep. {\bf 301}, 65 (1998)}.

\bibitem{BD2} T. Bodineau and B. Derrida, ``Distribution of current in nonequilibrium diffusive systems and phase transitions,Ó \href{http://journals.aps.org/pre/abstract/10.1103/PhysRevE.72.066110}{Phys. Rev. E {\bf 72}, 066110 (2005)}.

\bibitem{wePRE} C. P\'erez-Espigares, P. L. Garrido, and P. I. Hurtado,  ``Dynamical phase transition for current statistics in a simple driven diffusive system," \href{http://journals.aps.org/pre/abstract/10.1103/PhysRevE.87.032115}{Phys. Rev. E {\bf 87}, 032115 (2013)}.

\bibitem{weSSB} P. I. Hurtado and P. L. Garrido, ``Spontaneous symmetry breaking at the fluctuating level," \href{http://journals.aps.org/prl/abstract/10.1103/PhysRevLett.107.180601}{Phys. Rev. Lett. {\bf 107}, 180601 (2011)}.

\end{thebibliography}
\end{document}